\documentclass[fleqn,usenatbib,twocolumn]{mnras}

\usepackage{newtxtext,newtxmath}

\usepackage[T1]{fontenc}
\usepackage{ae,aecompl}

\usepackage{graphicx}	
\usepackage{amsmath}	
\usepackage{amssymb}	
\usepackage{color}
\usepackage{tabu}
\usepackage[usenames,dvipsnames,svgnames,table]{xcolor}
\hypersetup{
    citecolor = {gray},
    linkcolor = {black}
}
\usepackage[printwatermark]{xwatermark}

\newcommand{\pccc}{\rm cm^{-3}}	
\newcommand{\bn}{\begin{enumerate}}
\newcommand{\en}{\end{enumerate}}
\newcommand{\bi}{\begin{itemize}}
\newcommand{\ei}{\end{itemize}}

\newcommand{\Zsun}{Z_\odot}

\newcommand{\Msun}{M_\odot}

\newcommand{\Htwo}{{\rm H_2}}
\newcommand{\angstrom}{\textup{\AA}}
\newcommand{\RIII}{\mathcal{R}_{\rm III}}

\title[Legacy of early star formation]{Legacy of star formation in the pre-reionization universe}

\author[Jaacks et al.]{
Jason Jaacks,$^{1}$\thanks{E-mail: jaacks@astro.as.utexas.edu}
Steven L. Finkelstein$^{1}$ and
Volker Bromm$^{1}$
\\
$^{1}$Department of Astronomy, The University of Texas at Austin, Austin, TX 78712\\
}

\date{Accepted XXX. Received YYY; in original form ZZZ}

\pubyear{2017}

\begin{document}
\label{firstpage}
\pagerange{\pageref{firstpage}--\pageref{lastpage}}
\maketitle

\begin{abstract}
We utilize {\small GIZMO}, coupled with newly developed sub-grid models for Population~III (Pop~III) and Population~II (Pop~II), to study the legacy of star formation in the pre-reionization Universe. We find that the Pop~II star formation rate density (SFRD), produced in our simulation ($\sim 10^{-2}\ \Msun{\rm yr^{-1}\, Mpc^{-3}}$ at $z\simeq 10$), matches the total SFRD inferred from observations within a factor of $<2$ at $7\lesssim z \lesssim10$. The Pop~III SFRD, however, reaches a plateau at $\sim10^{-3}\ \Msun{\rm yr^{-1}\, Mpc^{-3}}$ by $z\approx10$, remaining largely unaffected by the presence of Pop~II feedback. At $z$=7.5, $\sim 20\%$ of Pop~III star formation occurs in isolated haloes which have never experienced any Pop~II star formation (i.e. primordial haloes). We predict that Pop~III-only galaxies exist at magnitudes $M_{\rm UV}\gtrsim -11$, beyond the limits for direct detection with the {\it James Webb Space Telescope (JWST)}. We assess that our stellar mass function (SMF) and UV luminosity function (UVLF) agree well with the observed low mass/faint-end behavior at $z=8$ and $10$. However, beyond the current limiting magnitudes, we find that both our SMF and UVLF demonstrate a deviation/turnover from the expected power-law slope ($M_{\rm UV,turn}= -13.4\pm1.1$ at $z$=10).   This could impact observational estimates of the true SFRD by a factor of $2 (10)$ when integrating to $M_{\rm UV} = -$12 ($-$8) at $z\sim 10$, depending on integration limits.  Our turnover correlates well with the transition from dark matter haloes dominated by molecular cooling to those dominated by atomic cooling, for a mass $M_{\rm halo}\approx 10^{8} \Msun$ at $z\simeq 10$.

 \end{abstract}



\section{Introduction}
\label{sec:intro}
The impending launch in $\sim 2020$ of the {\it James Webb Space Telescope (JWST)} promises the capability to collect photons from a yet unexplored epoch of cosmic evolution, expanding our view to a time when the Universe was $<400$~Myr old. The goal of the work presented here is to elucidate the legacy left behind by star formation in the pre-reionization Universe, at redshifts $z\gtrsim 7$, with state of the art cosmological simulations. We are thus addressing the crucial period of cosmic dawn, when the simple initial conditions of the very early Universe give way to an ever increasing complexity of cosmological structure. This is a very timely endeavor, given the powerful array of next-generation observational facilities that are currently being deployed or planned. 

For more than 25 years, the Hubble Space Telescope ({\it HST}) has been a cornerstone of modern high redshift ($z>5$) astronomy.  The 2009 installation of the near-infrared Wide Field Camera 3 (WFC3) instrument aboard {\it HST} opened up the $z>7$ Universe with a myriad of new photometric targets \citep[e.g.][]{Oesch.etal:09, Ouchi.etal:09, Bouwens.etal:10a, Finkelstein:10, Finkelstein:12c, Trenti.etal:11, Wilkins.etal:11a}. Large survey programs, such as CANDELS, BoRG, HUDF09/12, GOODS, CLASH, and the Hubble Frontier Fields (HFF), have discovered galaxies with luminosities spanning $\sim$10 magnitudes, and transformed our understanding of how the Universe has evolved over its first billion years.  

{\it JWST} will extend this dynamic range by $\sim3$ magnitudes ($m_{\rm AB}\approx32$ in a $\sim$200 hr blank field). Early release science programs, such as the Cosmic Evolution Early Release Science Survey\citep[CEERS: ][]{CEERS:18}, anticipate the detection of $\sim 50$ new $9\leq z \leq13$ galaxies in a $100$ arcmin$^2$ field with up to 10 residing at $z>11$. For context, there is currently only one spectroscopically confirmed galaxy at these redshifts from {\it HST} \citep[$z=11.1$; ][]{Oesch:16}. Equipped with this new information, we will begin to address several key questions which have been raised over the past decades of {\it HST} observations, such as: Is there a ``dearth'' of galaxies at $z>10$ as suggested by \citet{Oesch:18}? Does the UV luminosity function (UVLF) turn over, or maintain its power-law behaviour, at the faint-end?  Will we directly detect the first generation of stars, the so-called Population~III (Pop~III)?  If not, can we detect the legacy left behind by Pop~III processes, such as supernova (SN) explosions, metal-enriched absorption systems, or star clusters? Where should we look for these signatures?

The past decade has seen a slew of pioneering numerical work which has provided quantitative predictions for the upcoming {\it JWST} mission \citep[e.g.][]{Tornatore:07, Salvaterra.etal:11, Jaacks.etal:12a, Wise:14, Pallottini:15, Ma:17}. These studies have also begun to explore many of the fundamental questions mentioned above. For example, \citet{Wise:14} utilized the adaptive mesh refinement (AMR) code {\small ENZO} \citep{Bryan:14}, which is able to achieve excellent spatial resolution in select regions to study galaxy evolution at $z>7$. More specifically, this work focused on the impact of dwarf galaxies in low-mass haloes, $6.5\leq \log(M_{\rm halo}/\Msun)\leq 8.5$, on the ionizing photon budget in the pre-reionization epoch. \citet{Wise:14} present evidence that, even though below $M_{\rm UV}\sim -12$ the UVLF flattens, dwarf galaxies contribute $\sim30\%$ of the ionizing photons at $z$=6.

Work by \citet{Johnson:13} analyzed data from the First Billion Years (FiBY) simulation which utilized the smoothed particle hydrodynamics (SPH) code {\small GADGET} to focus on the evolution of the $z>6$ Universe. This investigation finds that significant Pop~III star formation continues down to at least $z\sim 6$, which, even though directly undetectable by {\it JWST}, gives hope for the possibility of observing a Pop~III spawned pair instability supernova (PISN). They calculate that at $z$=10, there could be up to one PISN visible per deg$^2$ per year \citep[see also][]{Scannapieco:05}.
Using the AMR code {\small RAMSES} to explore cosmic metal enrichment within the first galaxies, \citet{Pallottini:15} also find evidence for substantial Pop~III star formation, with star formation rate density (SFRD) $\sim10^{-3}\ \Msun\ {\rm yr^{-1}}\,{\rm Mpc}^{-3}$, down to $z$=6. Afterwards, Pop~III is rapidly quenched. They estimate that Pop~III constitutes $\sim 10\%$ of the total star formation at $z\sim7$. These authors also find that the mean total metallicity in their simulation volume from both Pop~III and Pop~II, $Z_{\rm Pop~III}+Z_{\rm Pop~II}$, does not cross the critical metallicity required for the Pop~III/II transition ($Z_{\rm crit}=10^{-4}\ \Zsun$) until $z\sim8.5$. The volume filling fraction of gas with $Z>Z_{\rm crit}$ is $\sim 10^{-3}$ at the same redshift, leaving a large reservoir of zero- or low-metallicity gas for ongoing Pop~III star formation.

The rapid pace of discovery implies that early predictions are not directly applicable to interpret recent frontier observations. The latter, however, can provide critical guidance in developing and testing cutting-edge numerical experiments. A case in point is the discovery of the previously mentioned $z=11.1$ galaxy \citep{Oesch:16}, identified by {\it HST} in the CANDELS/GOODS-N data, which to date provides our only direct constraints for the physical properties of $z\gtrsim 10$ galaxies. Specifically, this galaxy is $\sim 3$~times brighter than a typical $L_*$ galaxy at $z$=7, has an estimated stellar mass of $M_*=10^9\ \Msun$, and is forming stars at a rate of $\sim 25\ \Msun\ {\rm yr^{-1}}$. This rapid rate of star formation presents a challenge for standard galaxy formation models. 
\citet{Livermore:17} use data from the HFF to address the early predictions from \citet{Jaacks:13} and \citet{Wise:14}, claiming  a turnover or flattening of the UVLF at $z\geq 6$. Studying $\sim$170 lensed $z>6$ galaxies, \citet{Livermore:17} find steep faint-end slopes of $\alpha_{\rm UV}<-2$, but no evidence for any deviation from the power-law form down to limiting magnitudes of $M_{\rm UV}\sim-12.5$ at $z\sim6$, $M_{\rm UV}\sim-14.5$ at $z\sim7$, and $M_{\rm UV}\sim-15$ at $z\sim8$ \citep[see also ][]{Atek:15,Yue:16}. Conversely, considering Local Group descendants, \citet{MBK:15} argue that a flattening of the UVLF around the limits set by Livermore et al.\ is required to match Milky Way satellite galaxy number counts and reionization constraints at the limits probed by observations \citep[see also ][]{MBK:14}. A similar lensing program with {\it JWST} should enable exploration down to absolute UV magnitudes of $M_{\rm UV}\sim-11$, thus promising to greatly enhance our understanding of the UVLF.

Most recently, results from the Experiment to Detect the Global EoR Signature \citep[EDGES; ][]{Bowman:18} provide hints for the very onset of cosmic star formation. EDGES employs a low-frequency radio antenna, located in a radio quiet region of Western Australia, to detect a global absorption feature in the redshifted 21-cm hyperfine-structure signal of neutral hydrogen, seen against the cosmic microwave background (CMB). The signal is enabled by Lyman-$\alpha$ photons, produced in nebular emission around the first stars,  interacting with primordial hydrogen. The detection of a signal centered at 78~MHz suggests ongoing star formation already at a time when the Universe was a mere 180~million years old. This tantalizing observation offers our earliest constraint on star formation to date, nicely complementing the lower-$z$ data around the epoch of reionization.

In this work, we introduce a newly developed Pop~II star formation model in conjunction with our existing Pop~III legacy model \citep{Jaacks:17a}, to study star formation and metal-enrichment in the pre-reionization Universe ($z\gtrsim7$). This paper is organized as follows. In Section~\ref{sec:methods} we describe our numerical methodology, followed by the presentation of our results in Section~\ref{sec:results}. In Section~\ref{sec:predict}, we discuss key predictions for the upcoming {\it JWST} mission, and compare our work with previous studies in  Section~\ref{sec:disc}. We end in Section~\ref{sec:con} with our major conclusions.

\begin{table*} 
\caption{Simulation parameters used in this paper. The parameter $N_{\rm p}$ is the number of gas and dark matter particles; $m_{\rm DM}$ and $m_{\rm gas}$ are the particle masses of dark matter and gas; $\epsilon,h_{\rm sml}$ are the comoving gravitational softening length/hydrodynamical smoothing length (adaptive). 
}
\begingroup
\setlength{\tabcolsep}{6pt} 
\renewcommand{\arraystretch}{1.2} 
\begin{tabular}{cccccccc}
\hline
Run  	& Box size 	& $N_{\rm p}$ 		& $m_{\rm DM}$ 			& $m_{\rm gas}$ 		 & $\epsilon,h_{\rm sml}$  & Pop~III 	&	Pop~II\\ 
			& (Mpc $h^{-1}$) 	& (DM, Gas) 	& ($\Msun$)		& ($\Msun$) 			 & (kpc)  	&	model	&	model\\
\hline
N512L4 		& $4.0$ 	& $2 {\times} 512^{3}$ 	& $4.31{\times} 10^{4}$ 	& $9.64 {\times} 10^{3}$ & $0.45$ 	&P3L	& P2L \\
\hline
\end{tabular} 
\label{tbl:Sim}
\endgroup
\end{table*}

\section{Numerical Methodology}
\label{sec:methods}
For this work, we utilize a highly customized version of the publicly available next generation hydrodynamics/N-body code {\small GIZMO}, which employs a Lagrangian meshless finite-mass (MFM) method for solving the fluid equations. {\small GIZMO} offers improved numerical accuracy and efficiency when compared to previous generations of smoothed particle hydrodynamics (SPH) and adaptive mesh refinement (AMR) codes  (for a detailed method comparison, see \citealt{Hopkins:15}). In this section, we will describe our choice of parameters for the simulation volume, as well as our customized sub-grid physics models, developed for this work.

Our simulation volume, designed to approximately replicate a single pointing with the {\it JWST} at redshift $z\sim 10$, has a (comoving) box size of $4h^{-1}$~Mpc, and contains $512^3$ particles in both gas and dark matter. We will refer to this simulation run as N512L4 throughout, and provide full details of the set-up in Table~\ref{tbl:Sim}. Specifically, we adopt a $\Lambda$ cold dark matter ($\Lambda$CDM) cosmology, with parameters consistent with recent {\it Planck} results: $\Omega_m=0.315$, $\Omega_\Lambda=0.685$, $\Omega_b=0.047$, $\sigma_8=0.829$, and $H_0=67.74$~km~s$^{-1}$\,Mpc$^{-1}$ \citep{Planck:16}. Our initial conditions are generated at $z=250$, using the {\small MUSIC} initial conditions generator \citep{Hahn:11}.

Dark matter haloes are identified via a post processing 3D friends-of-friends (FOF) algorithm with a minimum particle requirement of $32$ and a linking length of $0.15$ times the inter-particle distance.  Gas particles and their respective properties (mass, temperature, metallicity, density, position) are then associated with each dark matter halo by searching within its virial radius.  Galaxy grouping utilizes a similar FOF algorithm and in all of the data presented in this work each of our galaxies has at least 32 total particles (gas + stars). Grouping and data extraction are aided by the {\it yt} \citep{Turk:yt} and {\it Caesar} \citep{pygr} software packages.

In the following subsections we will detail the custom sub-grid models which we have implemented into the public version of the {\small GIZMO} code base.  This work includes new prescriptions for Pop~II star formation and associated feedback, tracking metals created in Pop~II and Pop~III SN events independently, cooling via metal lines, and UV background heating.  For details regarding previously implemented models, regarding Pop~III star formation and feedback, primordial gas chemistry/cooling, Lyman-Werner (LW) photo-dissociation, and a stochastically sampled Pop~III initial mass function (IMF), we refer the reader to \citet{Jaacks:17b,Jaacks:17a}.

\subsection{Chemistry}
Primordial chemical abundances are calculated and tracked for 12 species (H, H$^+$, H$^-$, $\Htwo$, $\Htwo^+$, He, He$^+$, He$^{2+}$, D, D$^+$ HD, e$^-$), using methods detailed in \citet{Bromm:02} and \citet{Johnson:06}, in turn based on earlier work \citep{Cen:92,Galli:98}. As $\Htwo$ and hydrogen deuteride (HD) are the primary low-temperature coolants in primordial gas, it is critical to properly account for their formation and destruction.  Therefore, we also include $\Htwo$ photo-dissociation, as well as photo-detachment of ${\rm H}^{-}$ and $\Htwo^{+}$, via an external LW background in our chemical network, together with a prescription for self-shielding. In principle, lithium hydride (LiH) is another molecular coolant in low-temperature primordial gas. However, \citet{Liu:18} have shown that it is unimportant in most environments, including those encountered in the assembly pathway of the first galaxies. 

Heavy elements, which are the result of Pop~III and Pop~II star formation processes, are independently tracked in our simulation with passive metallicity scalar variables, and aggregated for each gas particle
\begin{equation}
Z_{\rm total} = Z_{\rm Pop~II} + Z_{\rm Pop~III}.
\end{equation}
Individual elemental abundances are then assigned in a post-processing procedure, accounting for the yield differences between Pop~II and Pop~III SN explosions.

\begin{figure}
\begin{center}
\includegraphics[scale=0.35	] {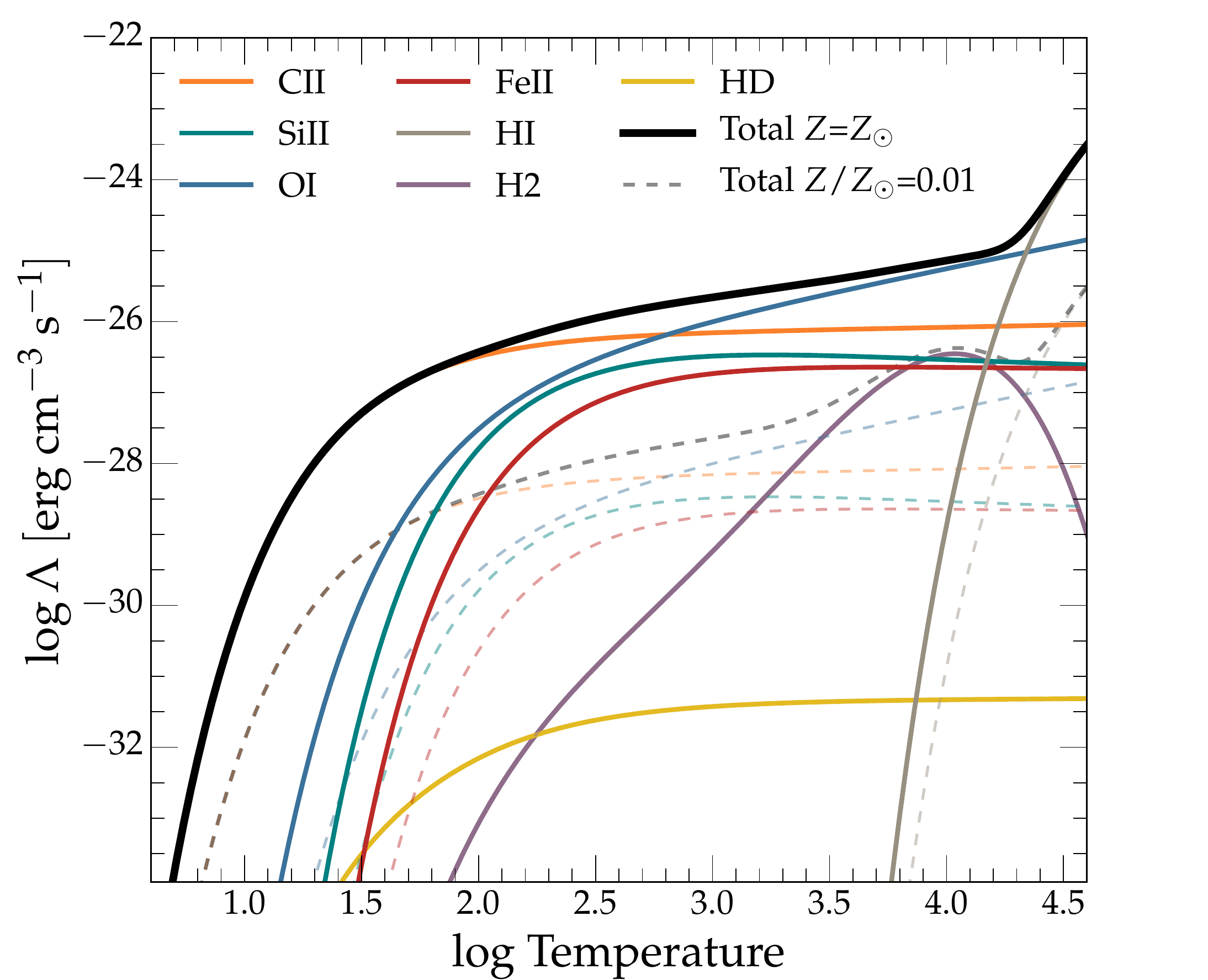}
\caption{Primary low temperature ($T\lesssim10^4$ K) cooling channels included in our simulations.  Fine structure metal cooling from \ion{C}{ii} (orange), \ion{O}{i} (blue), \ion{Si}{ii} (green), \ion{Fe}{ii} (red), following the detailed prescription presented in \citet{Maio:07}.  Molecular cooling with primordial contributions from $\Htwo$ (purple) and HD (yellow), as implemented in \citet{Galli:98}. Due to the large uncertainties in abundance patterns at high-$z$, we assume solar abundances according to \citet{Asplund:09}, which scale with total metallicity.  Solid lines represent $Z=\Zsun$ and the dashed lines are calculated for $Z/\Zsun=0.01$. For both metallicities, the fractional abundances of $\Htwo$ and HD are set to $10^{-5}$ and $10^{-7}$, respectively, with an electron fraction of $10^{-3}$.} 
\label{fig:cool}
\end{center}
\end{figure}

\subsection{Cooling and heating}
\label{subsec:coolingheating}
At each simulation time step, we evaluate the total cooling ($\Lambda$) and heating ($\Gamma$) rates for each particle in the simulation volume, to act as sink and source terms in the internal energy equation. 
In the following, we will describe the most important cooling and heating terms encountered in the formation of the first stars and galaxies. 

\subsubsection{Cooling}
\label{subsubsec:cooling}
In Figure~\ref{fig:cool}, we present the rates for the primary low-temperature cooling channels included in our simulations. Primordial cooling ($\Lambda_{\rm pri}$), in the absence of metals, is dominated by $\Htwo$ and HD, whereas gas enriched by preceding star formation cools via fine-structure metal lines ($\Lambda_{\rm fine}$), dominated by \ion{C}{ii}, \ion{O}{i}, \ion{Si}{ii} and \ion{Fe}{ii}. For simplicity, we assume that carbon, silicon, and iron are singly ionized by the soft UV (LW) background radiation field, and that oxygen is neutral due to its higher ionization potential \citep{Bromm:03b}. We adopt solar abundances according to \citet{Asplund:09}, scaling with total metallicity. However, there are large uncertainties associated with abundance patterns at high-$z$, including the possibility that they evolve with redshift \citep[e.g.][]{Cullen:16,Steidel:16}. We plan to explore the impact of evolving abundance patterns in future work.  The solid lines represent rates in gas enriched to $Z=\Zsun$, to provide a reference, and the dashed lines are calculated for $Z/\Zsun=0.01$, the typical metallicity in Pop~II star forming clouds. 

In our numerical implementation, cooling rates for \ion{C}{ii} and \ion{Si}{ii} are modeled as two-level systems utilizing
\begin{equation}
\Lambda_{2} = n_2A_{21}\Delta E_{21}\ {\rm erg \ s^{-1}cm^{-3}},
\end{equation}
where $n_2$ is the number density in the excited state, $A_{21}$ the Einstein coefficient for the spontaneous transition probability per unit time from levels $2 \rightarrow 1$, and $\Delta E_{21}$ the energy difference between the two levels.
\ion{O}{i} and \ion{Fe}{ii} are modeled as three-level systems with the contribution from each level summed
\begin{equation}
\Lambda_{3} = \sum_{i\geq 2}\sum_{1 \leq j<i} n_i A_{ij}\Delta E_{ij}\ {\rm erg \ s^{-1}cm^{-3}}.
\end{equation}
Our calculations follow closely details presented in \citet{Maio:07}, who employ data for excitation rates and energy levels given in \citet{Hollenbach:89} and \citet{Santoro:06}.

Our cooling channels for high temperature gas, $\Lambda_{\rm high}$, include atomic line cooling of \ion{H}{i} and \ion{He}{ii}, collisional ionization and recombination cooling (\ion{H}{i}, \ion{He}{i}, \ion{He}{ii}), as well as bremsstrahlung and inverse Compton cooling off the CMB \citep{Cen:92,Greif:07}.
It is worth noting that, despite large abundances even at high redshift, we do not include CO molecular cooling. This is justified by the fact that cooling via rotational transitions of CO is largest at temperatures which are less than the redshift-dependent CMB floor, imposed in our simulations (see \citealt{Omukai:10} for details regarding CO cooling).

\subsubsection{Heating}
\label{subsubsec:heating}
In addition to the LW background heating implemented in \cite{Jaacks:17a}, we here include two new heating terms, due to an external ionizing UV background ($\Gamma_{\rm uvb}$) and to local photoelectric absorption  ($\Gamma_{\rm pe}$).

We model the UV background (UVB) heating by including the term 
\begin{equation}
\Gamma_{\rm uvb} = e^{-\tau}\zeta_{\rm ion}(z) n_{\rm HI} \left< \psi \right> k T_c\ {\rm erg \ s^{-1}cm^{-3}}, 
\label{eq:uvb}
\end{equation}
where $n_{\rm HI}$ is the number density of neutral hydrogen, and $\left<\psi\right> k T_c$ the mean photon energy produced by a given stellar population. The color temperature is chosen to be consistent with a young stellar population dominated by O and B stars, $T_c=20000$~K, corresponding to a mean value of $\left<\psi\right>=1.38$, obtained from \citet{Spitzer:78}.  $\zeta_{\rm ion}(z)$ is the redshift dependent photo-ionization rate calculated in \cite{FG:09}, and updated in 2011. For reference, $\zeta_{\rm ion}(z=6)\approx 2.80\times10^{-13}\ {\rm s^{-1}}$.  $\Gamma_{\rm uvb}$ is calculated at each time step and applied to all gas particles in the simulation volume. 

Included in Equation~\ref{eq:uvb} is the term $e^{-\tau}$, which allows us to take into account the ability for high density gas to self-shield against the UVB. Here, the effective optical depth to ionizing photons is evaluated as
\begin{equation}
\tau \simeq \sigma_{\rm th} N_{\rm HI} \simeq  \sigma_{\rm th} n_{\rm HI} L_{\rm char},
\end{equation}
where $\sigma_{\rm th}\simeq6.3\times10^{-18}\ {\rm  cm^{2}}$ is the hydrogen
photo-ionization cross-section at the threshold (13.6~eV), $N_{\rm HI}$ the local neutral hydrogen column density, and $L_{\rm char}$ the characteristic size of the system.  With this term, high-density star forming regions are able to shield against the effects of the UV background radiation. We note that it is standard numerical practice to estimate $L_{\rm char}$ using the hydrodynamical smoothing length of a gas particle (see \citealt{CSS:12} for comparison of different methods to determine $L_{\rm char}$).  However, very low-density gas particles in the IGM will have very large smoothing lengths, thus artificially boosting their shielding ability. Therefore, we fix $L_{\rm char}=1$~kpc in order to avoid unphysically large IGM opacity.  Note, both our LW and UVB heating terms are applied homogeneously in our computational volume and do not take into account spatial clustering.

To estimate the local photoelectric heating, $\Gamma_{\rm pe}$, due to stellar populations in our simulation volume, we consider each star particle as a simple stellar population, which we in turn regard as being spatially co-located. We thus model the combined emission from a stellar cluster as a point source, which is justified given that we do not resolve the size of a typical cluster. The volumetric photoelectric heating rate for each stellar population is then calculated as
\begin{equation}
\Gamma_{\rm pe}({\rm H}\rightarrow {\rm H}^+) = \alpha_{B} n_{\rm H}^2 \left< \psi \right> k T_c\ {\rm erg \ s^{-1}cm^{-3}}, 
\end{equation}
where $\alpha_B=2.59 \times 10^{-13}\ {\rm cm^3s^{-1}}$ is the case-B recombination rate coefficient, $n_{\rm H}$ the hydrogen number density, and  $\left<\psi\right> k T_c$ the mean photo-electron energy, released in a typical interaction. This heating term is applied at each time step to particles which are found within the Str$\ddot{\rm o}$mgren radius of a Pop~II star formation event, for the estimated lifetime of OB stars ($\sim 10$ Myr).  Note that the simulation timestep during this stage is $\sim10^5$~yr, such that this effect is temporally well resolved. We will describe the approach to calculate the Str$\ddot{\rm o}$mgren radius, $R_{\rm Stromgren}$, in Section~\ref{sec:feedback}.

\subsection{Pop III legacy star formation}
Our  Pop III ``Legacy'' (P3L) star formation model allows us to essentially ``paint'' fully formed SN blast waves onto our simulation boxes, centered on Pop~III star forming regions, with physical properties calibrated to high-resolution simulations \citep[e.g.,][]{Greif:07,Ritter:12}, and to well-known analytic solutions \citep{Sedov:59,Taylor:50}. Each star formation event has a stellar population, which is randomly drawn from a given IMF,  here taken to be approximately flat with a low-mass turn over at $\sim 5 \Msun$.  Our adopted IMF is consistent with results from high-resolution Pop~III star formation simulations \citep[e.g.][]{Greif:11, Stacy:13}. This allows each star forming event to exhibit a unique feedback signature, in terms of explosion energy and nucleosynthetic yield, because stars with different masses end their lives differently, as type~II SN, black hole, or pair-instability SN. The feedback bubble radius, metallicity, thermal energy, and ionization are directly calculated for each individual stellar population. Thereafter, the enriched gas is advected with the local hydrodynamical flow. This approach enables our star formation model to have both time and spatial dependence, and to independently trace the Pop~III and Pop~II origins of the aggregate metallicity.  Full details of the P3L model can be found in \cite{Jaacks:17a}.

\subsection{Pop~II star formation}
\subsubsection{Formation criteria}
\label{sec:sf_cri}

Pop~II star formation is triggered when a preset threshold density, $n_{\rm th}=100$~cm$^{-3}$, is reached for a gas particle with $T\leq 10^3$ K and $Z/\Zsun >10^{-4}$. The latter represents the critical metallicity, required to transition from Pop~III star formation to Pop II \citep[e.g.][]{CSS:10,Schneider:12}. This density is adopted due to the mass/spatial resolution limitations of our simulation volumes, which are only able to resolve pre-stellar clump scale objects, and not the individual star forming cores contained within.  However, $n_{\rm th}$ is physically representative of densities which are observed in local star forming regions \citep[e.g.][]{Bergin:07,McKee:07}, and sufficient to ensure that the cooling processes modeled here are efficient. Therefore, we are able to identify regions with the physical conditions necessary for runaway gravitational collapse.

\subsubsection{Stellar population}
\label{sec:imf}
\begin{figure}
\begin{center}
\includegraphics[scale=0.35] {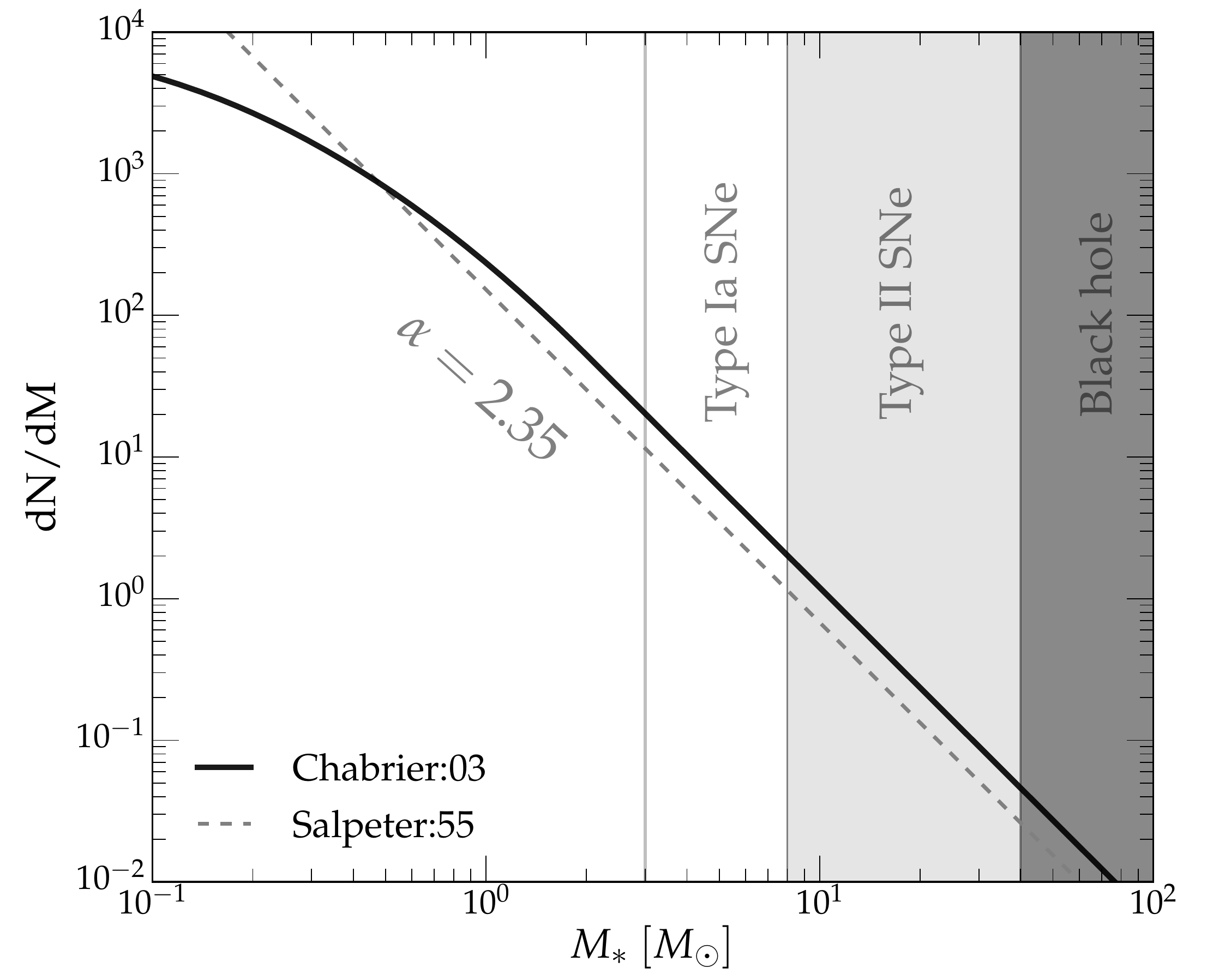}
\caption{The \citet{Chabrier:03} IMF adopted in this work, normalized to the total mass of our SSP. With the shaded regions, we indicate mass ranges associated with different stellar fates, i.e. type~Ia SNe, type~II SNe, direct-collapse black holes \citep[][]{Heger:02}. For comparison, we show a \citet{Salpeter:55} IMF, normalized to the same total stellar mass (dashed line). } 
\label{fig:imf}
\end{center}
\end{figure}

\begin{table} 
\begin{center}
\caption{Key SSP characteristics. Stellar mass ($M_*$), total number of stars ($ N_*$), total number of OB stars ($N_{\rm O+B}$), number of Type~II SNe ($N_{\rm SNe}$), total mass of Type~II SN progenitor stars ($M_{*,\rm SNe}$), and energy injected per SN ($E_{\rm SN}$).  The values below have been calculated over a mass range of $0.08-100\ \Msun$ for a \citet{Chabrier:03} IMF.}
\setlength{\tabcolsep}{6pt} 
\renewcommand{\arraystretch}{1.2} 
\begin{tabular}{ccccccc}
\hline
  $\eta_*$ & $M_*$ & $N_*$ & $N_{\rm O+B}$ & $N_{\rm SNe}$ & $M_{*,\rm SNe}$ & $E_{\rm SN}$ \\
	     	            &           $[\Msun]$&  &   &  &    $[\Msun]$ &  [erg]\\
\hline
  $0.10$  & $964$	& $1323$  & $44$  & $11$ &  $160$ & $10^{51}$\\
\hline
\end{tabular}
\label{tbl:imf}
\end{center}
\end{table}

Once a star formation event is triggered, a star particle is instantaneously created with a stellar mass of $\eta_* \times m_{\rm gas}=964 \Msun$, where $\eta_*=0.10$ is the star formation efficiency on the proto-stellar clump scale.
In our prescription, the total stellar mass of a single formation event is fixed and has no local environmental dependency, e.g. on cloud mass or density.  This precludes the formation of large star clusters.  However, in a given halo, multiple dense clumps can arise that are not spatially co-located. Therefore, any given halo will be able to spawn a number of independent clusters. Once a gas particle qualifies for star formation, it exists as a point mass with gravitational interactions only, and will not be able to form any future stars. The remaining gas mass from the spawning SPH particle, i.e. $(1-\eta_*) m_{\rm gas}$, is considered to be part of the local stellar system and is removed from the hydrodynamic calculations, but it still contributes to the gravitational force.
Since the ``locked-in'' gas mass is permanently unavailable for hydrodynamic interactions, and the star particle mass is effectively $\sim$10 times higher than the actual stellar mass in terms of the gravity solver, the detailed galaxy morphology in the vicinity of the star-formation sites is not accurately captured in our simulations. On the meso-scales of the bound structures and the IGM, which are the main focus of this study, this caveat has likely only a minimal impact on the results presented below, however.

We treat each star particle as a simple stellar population (SSP) with an IMF, which is taken to be \citet{Chabrier:03} 
\begin{equation}
    \xi(m)=\frac{dn}{d\log m}\propto
    \begin{cases}
       m_0^{-x}\exp \left[ {-\frac{(\log m - \log m_c)^2}{2\sigma^2}} \right], &  m\leq m_0 \\
      m^{-x}, &  m>m_0
    \end{cases}
\end{equation}
with a slope of $x=1.35$, $m_c=0.18$, $m_0=2.0$ (both in units of solar mass), and $\sigma = 0.579$, over a mass range of $0.08-100\ \Msun$ \citep{Chabrier:14}. For $m>m_0$, the \citet{Chabrier:03} IMF is identical to the \citet{Salpeter:55} one.  The exponential cutoff at $m<m_0$ results in a stellar population, which is slightly less bottom heavy. 
In Figure~\ref{fig:imf}, we show both the \citet{Chabrier:03} and \citet{Salpeter:55} IMF for comparison, whereas in Table~\ref{tbl:imf}, we present the relative number of stars for each evolutionary fate. The occurrence of each feedback event will determine the total energy and metal enrichment, returned to the surrounding medium.

\subsection{Pop~II legacy feedback}
\label{sec:feedback}
As with our P3L model, our primary concern is the metal enrichment legacy, resulting from Pop~II star formation. Stellar feedback processes, such as the production of ionizing photons and metals, play a critical role in the regulation of ongoing star formation. Previous generations of numerical simulations have shown that without stellar feedback, the Pop~II star formation rate density is unphysically large \citep[e.g.][]{Springel:03}.  Therefore, we include a multi-component stellar feedback prescription, where we consider both photo-electric heating from young stellar populations, as well as the thermal energy input and metal enrichment from Type~II SNe. Due to the average delay time of $\sim 1$~Gyr between progenitor formation to SN explosion \citep{Maoz:12}, we here do not consider Type~Ia SNe, as we are only concerned with redshifts $z>6$.

\begin{figure}
\begin{center}
\includegraphics[scale=0.3] {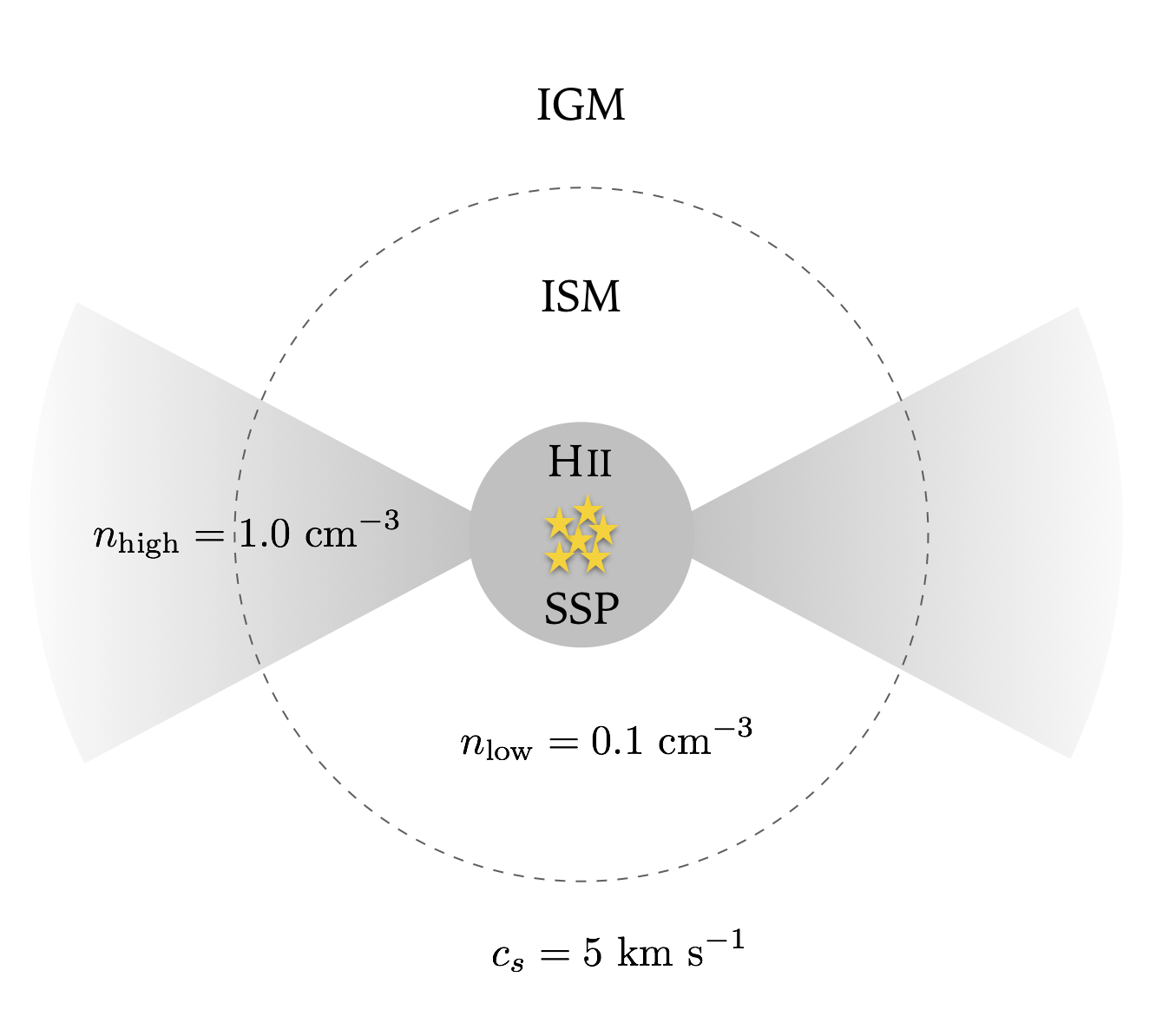}
\caption{Conceptual illustration of a typical star forming region in our Pop~II star formation model (not to scale). We indicate representative physical properties, assumed in each region for the post starburst, pre-SN phase.} 
\label{fig:region}
\end{center}
\end{figure}

\subsubsection{Thermal energy input}

As discussed in Section~\ref{subsubsec:heating}, $\Gamma_{\rm pe}$ is folded into our total heating term, $\Gamma_{\rm tot}$, and is applied to each gas particle which falls within the Str$\ddot{\rm o}$mgren radius associated with the stellar population
\begin{equation}
R_{\rm Str\ddot{o}mgren}= \left(\frac{3 \dot{Q}_{\rm ion}N_{*,\rm O+B}}{4\pi n_{\rm H_I}^2 \alpha_B}\right)^{1/3}\ \approx 240\ {\rm pc}.
\label{eq:rion}
\end{equation}
Here $\dot{Q}_{\rm ion}\approx10^{49}\ {\rm s^{-1}}$ is the average number of ionizing photons produced per second and per OB star, $n_{\rm HI}=1.0\ \pccc$ the neutral hydrogen number density in the interstellar medium (ISM) of a typical host halo, and $N_{*,\rm O+B}$ the number of OB stars producing ionizing photons.  The adoption of $n_{\rm HI} = 1\ {\rm cm}^{-3}$ is intended to approximate the typical physical conditions found in the post-starburst ISM, after the gas has hydrodynamically responded to the prolonged photo-heating from the central star cluster. $\Gamma_{\rm pe}$ is applied to each particle initially found within $R_{\rm Str\ddot{o}mgren}$, for the duration of the OB stars lifetime of $\sim 10$~Myr. It is important to note that we do not decouple from the hydrodynamics at any point in the SF routine.  Therefore, any increase in temperature (internal energy) will have a direct effect on the density of the surrounding gas.

\subsubsection{Metal enrichment}
Once the photo-electric heating phase has ended, our feedback model ``detonates`` a cumulative Type~II SN event, centered on the star particle, with a total energy  of $E_{\rm tot,Pop~II}=N_{\rm SNe}\times E_{\rm SN}$, where we assume instantaneous explosion of all contributing stars for simplicity (see Table~\ref{tbl:imf} for values). The total SN energy is then used as input for an expanding shell calculation, similar to what was done for Pop~III in \citet{Jaacks:17a}. The result in each case is an expression for the final radius of a spherical shell, where the SN blast wave stops expanding. For this work, we modify our model assumptions slightly, to better reflect the physical environments in which Pop~II stars form, such as higher mass host haloes and larger central gas densities. More specifically, we take the surrounding ISM to be at a density of $n=1.0\ \pccc$, as opposed to the previously adopted $n=0.1\ \pccc$.  Using this prescription, we find that $r_{\rm shell,high}\propto E^{0.35}_{\rm tot,Pop~II}\approx 900\ {\rm pc}$.  Further, due to the inhomogeneity of star forming clumps and the propensity for shells to seek out low density expansion channels (voids), we also calculate the shell radius for propagation through a low-density ISM region,  where $n=0.1\ \pccc$, resulting in $r_{\rm shell,low}\propto E^{0.38}_{\rm tot,Pop~II}\approx 1.2\ {\rm kpc}$. In Figure~\ref{fig:region}, we illustrate the physical assumptions in our Pop~II star formation model (not to scale). For the final enrichment radius, we adopt the approximate mean of the above values, $\overline{r}_{\rm shell}= 1.0\ {\rm kpc}$. Our parameters are consistent with recent high-resolution simulations of Pop~II star forming galaxies at high redshifts (see fig.~8 in \citealt{MJ:15}).

Metals are then equally distributed to each of the gas particles which are identified to be within $\overline{r}_{\rm shell}$, in accordance with $M_Z=M_{*,\rm SNe}\times y_Z\approx 16.0\ \Msun$, for an effective Pop~II SN yield of $y_Z=0.10$ \citep{Nomoto:13}. The metals are subsequently allowed to simply advect with the local cosmic flow, as a component of the original gas particle. The total metallicity of each gas particle is continuously updated as the aggregate of the contribution of metals from Pop~III and Pop~II enrichment.  

In addition, we also include a thermal component which heats the gas contained within $\overline{r}_{\rm final}$ to $T_{i}\approx T_{\rm IGM,Hot}\approx 10^4\ {\rm K}$ at the end of the OB stars lifetime. This is done to approximate the thermal impact of the expanding shell at the time when it has reached its final, stalling radius. It should be noted that we do not impart energy in the form of a momentum kick to particles, as our resolution does not allow us to properly follow the internal dynamics of the expanding shell. Our sub-grid prescriptions for both Pop~III and Pop~II are intended to reflect the long-term impact on the ISM and IGM, specifically the metal enrichment and boost in ionization, based on results from sophisticated, high-resolution ab initio simulations, which are able to self consistently follow the expansion of the radiation I-front, and the expanding SN blast wave through the ISM and into the IGM \citep[e.g.][]{Greif:07,Ritter:12,MJ:14,MJ:15}.

As pointed out in Section~\ref{sec:imf}, during a star formation event only 10\% of a gas particle is converted into stars. However, the remaining 90\% of the gas mass remains "locked" into the new star particle, which is no longer included in the hydrodynamical update. Initially, this assumption is approximately valid, but it is clear that eventually, after a `recovery time' of a few 10~Myr \citep[e.g.][]{MJ:14}, the locked-up gas would be returned into the star-forming ISM. Thus, our results need to be interpreted with this caveat in mind, when considering the later stages of our simulation.    

To explore the impact of the "locked-in" gas on our results, we examine the ratio of total "free" gas found in a given galaxy to the total ``locked`` gas mass ($ M_{\rm g,free}/M_{\rm g,locked}$).  At $z=8$, on average, galaxies in our simulation volume exhibit $\rm M_{g,free}/M_{g,locked}\gtrsim 20$.  For high density gas with $n>100\ \pccc$, eligible for potential star formation, the ratio still is $\rm M_{g,free}(n>100\ \pccc)/M_{g,locked}\gtrsim 10$.  This indicates that, at least to this point in the simulation, the amount of ``locked-in`` gas is not significant in terms of available gas to form stars.  
Similarly, removing gas from the hydrodynamics could also impact our metallicity estimates, given that $Z=M_{\rm metal}/M_{\rm gas}$. If we explore the most extreme scenario, considering the metallicity of a gas particle immediately adjacent to a star particle, then the ``true'' value would be $Z=M_{\rm metal}/(M_{\rm gas}+0.9M_{\rm gas})$. Even in this case, our metallicity estimates would thus only be too high by a factor of about two. However, in our treatment the SN metal ejecta are distributed over a large fraction of the gas particles in the bound structure (i.e. the halo or galaxy), or across the entire simulation volume for global averages.  Therefore, we do not expect the ``locked`` gas to have a significant impact on the results presented in this work

\section{Results}
\label{sec:results}

\subsection{Global properties}
\label{subsec:global}
\subsubsection{Star formation rate density}
\label{subsubsec:sfrd}

\begin{figure*}
\begin{center}
\includegraphics[scale=0.6] {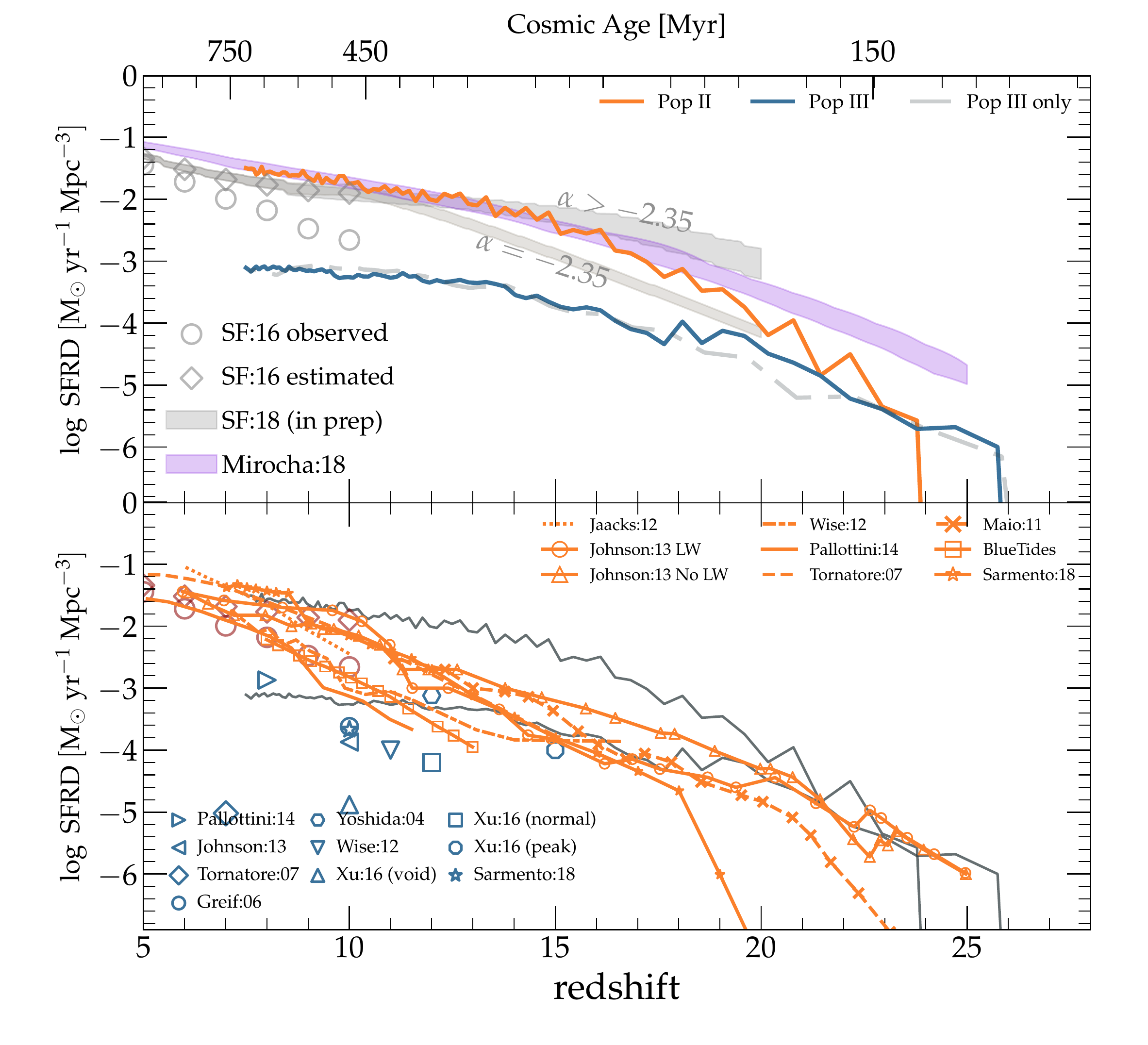}
\caption{{\it Top:}  Time averaged star formation rate density ($10$ Myr time bins) as a function of redshift, or cosmic age, for Pop~II (solid orange) and Pop~III (solid blue). For comparison, we reproduce the SFRD from our previous Pop~III-only work \citep{Jaacks:17a}, where we modeled the radiative feedback from Pop~II in an approximate way (gray dashed line). We also show observations for the Pop~II SFRD, derived from the reference UVLF (diamonds, circles) discussed in \citet{Finkelstein:16}, which combines frontier observations at $z\geq 5$ from published studies \citep[e.g.][]{McLure:09,McLure:13,Oesch:13,Oesch:14,Bowler:14,Schmidt:14,Bouwens:15a,Bouwens:15c,Finkelstein:15,McLeod:15}. Furthermore, we include the recent model estimate from \citet{Mirocha:18}, based on the 21cm absorption feature detected by EDGES \citep{Bowman:18}. {\it Bottom:} Comparison between this work, shown in dark gray solid lines, with previous numerical studies. Direct comparisons for Pop~III can be made between the lower solid gray line and the open blue symbols \citep{Yoshida:04, Greif:06, Tornatore:07, Wise:12, Johnson:13, Pallottini:15, Xu:16b, Sarmento:18}.  Direct comparisons for Pop~II can be made between the upper solid gray line and the various orange lines \citep{Greif:06, Tornatore:07, Maio:10, Jaacks.etal:12a, Johnson:13, Pallottini:15, Feng:16, Sarmento:18}. It should be noted that we are comparing our model estimates to extrapolations of current frontier observations throughout this work and not direct observables.
} 
\label{fig:sfrd}
\end{center}
\end{figure*}

As this work heavily depends on the star formation routines (P2L \& P3L), we first examine the star formation rate density (SFRD), produced over cosmic time in our simulation. In Figure~\ref{fig:sfrd}, we present the SFRD evolution for both Pop~III and Pop~II, covering the entire simulation volume (solid blue and orange lines, respectively). We witness the onset of Pop~III star formation occurring at $z\sim 26$ (cosmic age $\sim120$~Myr), followed promptly by a burst of Pop~II star formation at $z\sim 24$ (cosmic age $\sim 135$~Myr).  The delay time of $\sim 15$~Myr between the initial Pop~III activity and the subsequent round of Pop~II star formation is consistent with results from ultra-high resolution simulations, where the recovery timescale for second-generation star formation is estimated \citet{MJ:14}. Pop~II and Pop~III stars form at a fairly comparable rate until $z\sim 15$, whereas afterwards Pop~II star formation dominates by more than an order of magnitude over the remainder of our simulation run.  

It is interesting to note that the Pop~III SFRD, when Pop~II star formation is self-consistently included, deviates only minimally from our previous Pop~III-only simulations in \citet{Jaacks:17a}. This result suggests that Pop~III star formation is largely decoupled from the effects of Pop~II star formation and feedback, likely due to the sequence, where Pop~III locally always precedes Pop~II, and the biased environment of formation. In a way, Pop~III acts as a `pathfinder' for all subsequent star formation, and as long as there is available primordial gas at high density, the initial Pop~III star formation will follow this near-universal path. We examine this question further in Section~\ref{subsec:occur}.

In the top panel, we compare our simulation results to observational estimates for the $z=6-10$ SFRD, which is derived by integrating the observed UVLF and applying a conversion from luminosity to stellar mass density \citep[e.g. ][]{Kennicutt:98,Madau:14}. For direct comparison, we include observational constraints for the $z\lesssim 10$ Pop~II SFRD \citep[circles, diamonds;][]{Finkelstein:16}. The data points are derived from a "consensus" UVLF, which combines frontier observations from various published studies at $z\gtrsim 5$ (see Fig.~\ref{fig:sfrd} for references). Specifically, the circles are obtained by integrating the UVLF down to an observational limit of $M_{\rm UV}\sim -17$, whereas the diamonds are integrated down to a theoretical limit of $M_{\rm UV}\sim -13$. The latter aims to account for systems beyond current telescope capabilities. We find excellent agreement, within factors of $<2$, at $z\lesssim 10$ with the empirically estimated total SFRD.  Note, we are comparing our model estimates to extrapolations of current frontier observations throughout this work and not direct observables.

At redshifts $z=10-15$, our results are consistent with extrapolations for the total SFRD, when the faint-end UVLF power-law slope is allowed to evolve to values steeper than $\alpha_{\rm UV}=-2.35$ at $z\gtrsim 9$. At even higher redshifts, our results are similarly consistent with the empirically-based models of Finkelstein et al. (2018, in prep.), which apply physically motivated star formation cut-offs to explore reionization scenarios with low UV escape fractions. Specifically, they explore scenarios where the slope continues to evolve at $z >10$, and one where it remains fixed to its $z = 10$ value towards higher redshifts. At $z = 10-15$, our results are consistent with the evolving faint-end slope extrapolation, falling to slightly lower values beyond that, between the evolving and fixed faint-end slope empirical constraints. While considering an evolving UVLF is supported both by numerical works \citep{Trenti.etal:10, Jaacks.etal:12a} and observations \citep{Bouwens.etal:12b, Finkelstein:15}, the true nature of the faint end of the UVLF at $z\gtrsim 10$ is highly uncertain.  Therefore, we view the observation-based estimates as upper and lower bounds for the $z\gtrsim 10$ SFRD, with our model prediction ranging in between. We explore this further in Section~\ref{subsubsec:uvlf}, where we discuss our simulated UVLF.

Currently, direct observations of the $z>10$ SFRD are lacking, until the next generation of ground- and space-based telescopes. Therefore, we must turn to previous numerical simulations to provide additional validation for our Legacy star formation approach. We have shown in \citet{Jaacks:17a} that our P3L model produces results which are consistent with a wide range of previous numerical estimates for the Pop~III SFRD \citep[e.g.][]{Yoshida:04,Greif:06, Tornatore:07, Wise:12, Johnson:13, Pallottini:15, Xu:16b}. Results from these previous studies are shown as the blue symbols in the bottom panel of Figure~\ref{fig:sfrd}. Also shown in the bottom panel of Figure~\ref{fig:sfrd} is a direct comparison between our prediction for the Pop~II SFRD (solid gray line) and those produced by other numerical experiments \citep[various orange lines;][]{Greif:06, Tornatore:07, Maio:10, Jaacks.etal:12a, Wise:12, Johnson:13, Pallottini:15, Feng:16}. 

We find that, while our simulations agree very well with the direct observations and observation-based estimates at $z=6-10$, we deviate from most of the presented numerical works at $z\gtrsim 10$ by approximately an order of magnitude.  We discuss the possible causes for this discrepancy in Section~\ref{subsec:dev}. A notable exception to the this trend are the results presented in \citet{Mirocha:18}, who take a semi-analytical approach to galaxy evolution in order to determine if the observed galaxy population can account for the 21-cm EDGES observations. The purple shaded region in the top panel of Figure~\ref{fig:sfrd} represents the SFRD recovered from their model, with a star formation efficiency (SFE) calibrated to produce the observed $\sim 78$ MHz EDGES signal. Our simulated Pop~II SFRD agrees very well with the \citet{Mirocha:18} model out to $z\sim 20$, where we begin to see a deviation to lower values. This may be due to the resolution limits in our simulation volume. We should caution that the EDGES results \citep{Bowman:18}, while extremely exciting, need to be confirmed by other instruments and subjected to further cross-checks in the future. Any conclusions must, therefore, be considered as preliminary. 

\begin{figure}
\begin{center}
\includegraphics[scale=0.52] {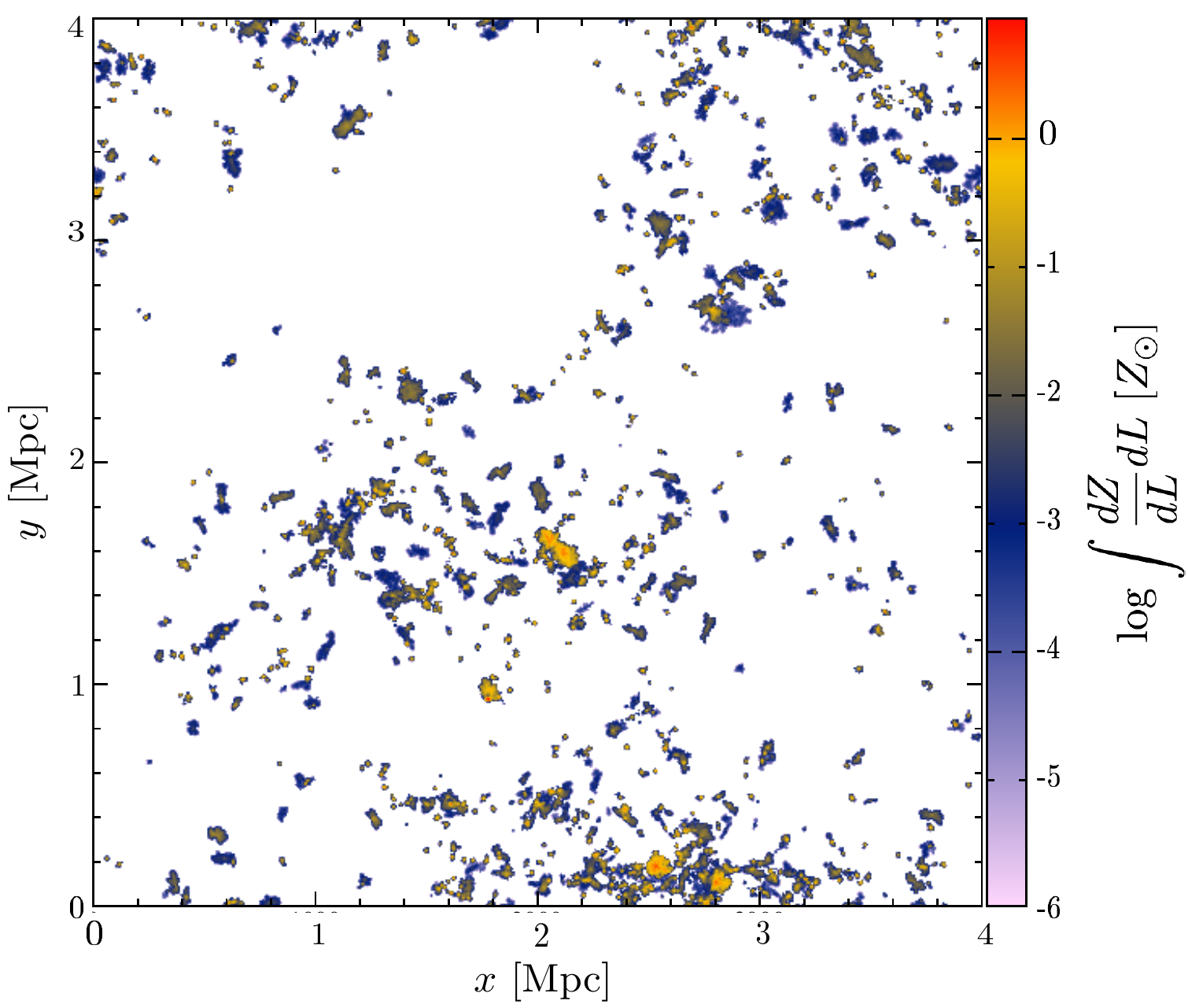}
\caption{Metal-enrichment legacy in the early IGM. Shown is the total metallicity (Pop~II+Pop~III) at $z=8$ for the entire simulation volume, in projection by integrating along the $z$-axis. All plots with projected metallicity, presented in this work, are rendered using {\small SPLASH} \citep{Price:07}. As can be seen, on the scale of the general IGM, a substantial fraction of the volume remains chemically pristine at the epoch of reionization.
}
\label{fig:viz_full}
\end{center}
\end{figure}

\begin{figure*}
\begin{center}
\includegraphics[scale=0.65] {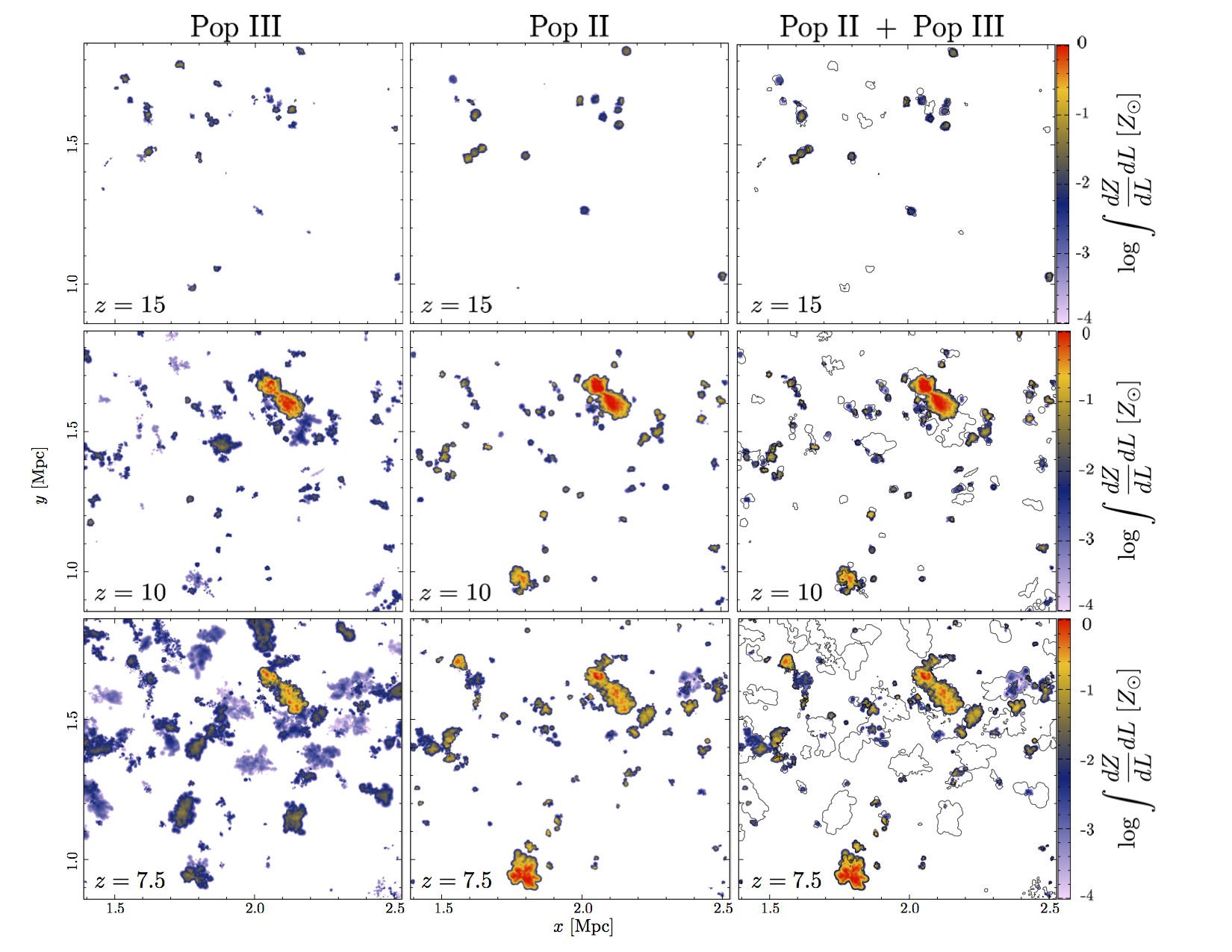}
\caption{Metallicity projection plots for a subsection of the simulation volume for Pop~III (left column), Pop~II (center column), and Pop~II+Pop~III (right column), shown at $z=15, 10, 7.5$ (top, center, bottom). In the right column, $Z_{\rm Pop~II}$ is shown by the color gradient, while $Z_{\rm Pop~III}$ is represented by a simple contour line. In this rendering, we are able to illustrate the regions, which are only enriched by Pop~III metals, clearly distinguishing them from those, where both Pop~II and Pop~III contribute to the enrichment. The `pathfinder' nature of Pop~III enrichment is evident, as it sets the stage for the subsequent Pop~II contribution.
} 
\label{fig:proj}
\end{center}
\end{figure*}

\subsubsection{Multi-component enrichment}
\label{subsubsec:enrich}

The primary legacy left behind by the formation and death of Pop~III and second-generation Pop~II stars is the enrichment with heavy chemical elements, which they impart to their environment. In Figure~\ref{fig:viz_full}, we present the total metallicity, integrated along the $z$-axis for our entire simulation volume at $z=7.5$. This rendition gives a qualitative understanding of the extent of metal enrichment, and the fraction of the cosmic volume impacted. In Section~\ref{subsubsec:vff}, we will quantify the volume filling fraction over cosmic time, thus providing a detailed analysis. However, it is clearly evident that towards the end of reionization, a substantial fraction of the IGM volume remains primordial, devoid of any metals. Those chemically pristine regions may provide `relics' of the end of the cosmic dark ages, to be probed with extremely deep absorption spectroscopy in the local Universe. 

In Figure~\ref{fig:proj}, we present the same projected metallicity along the $z$-axis, but now for only a sub-section of our simulation volume at $z$=15, 10, 7.5 (top, bottom, middle rows, respectively). Furthermore, we separate the contributions from Pop~III and Pop~II metals into separate columns (left and center). Finally, in the right-most column, we combine the Pop~II and Pop~III metal enrichment, such that the latter is shown with the single black contour lines, and the former with the color gradient patches.  With this presentation, we are able to qualitatively show both the increase in metallicity with cosmic time, and the independent contributions from each component to the total cosmic metallicity. Evidently, in this region of our simulation volume, there remain large regions, enriched by Pop~III only at $z$=7.5.  This result is consistent with the fact that the Pop~III SFRD is still high at $z$=7.5, and has yet to be terminated.  This also suggests that Pop~III star formation is continuing to occur in pristine haloes at these redshifts.

\subsubsection{Stellar mass function}
\label{subsubsec:smf}

\begin{figure}
\begin{center}
\includegraphics[scale=0.40] {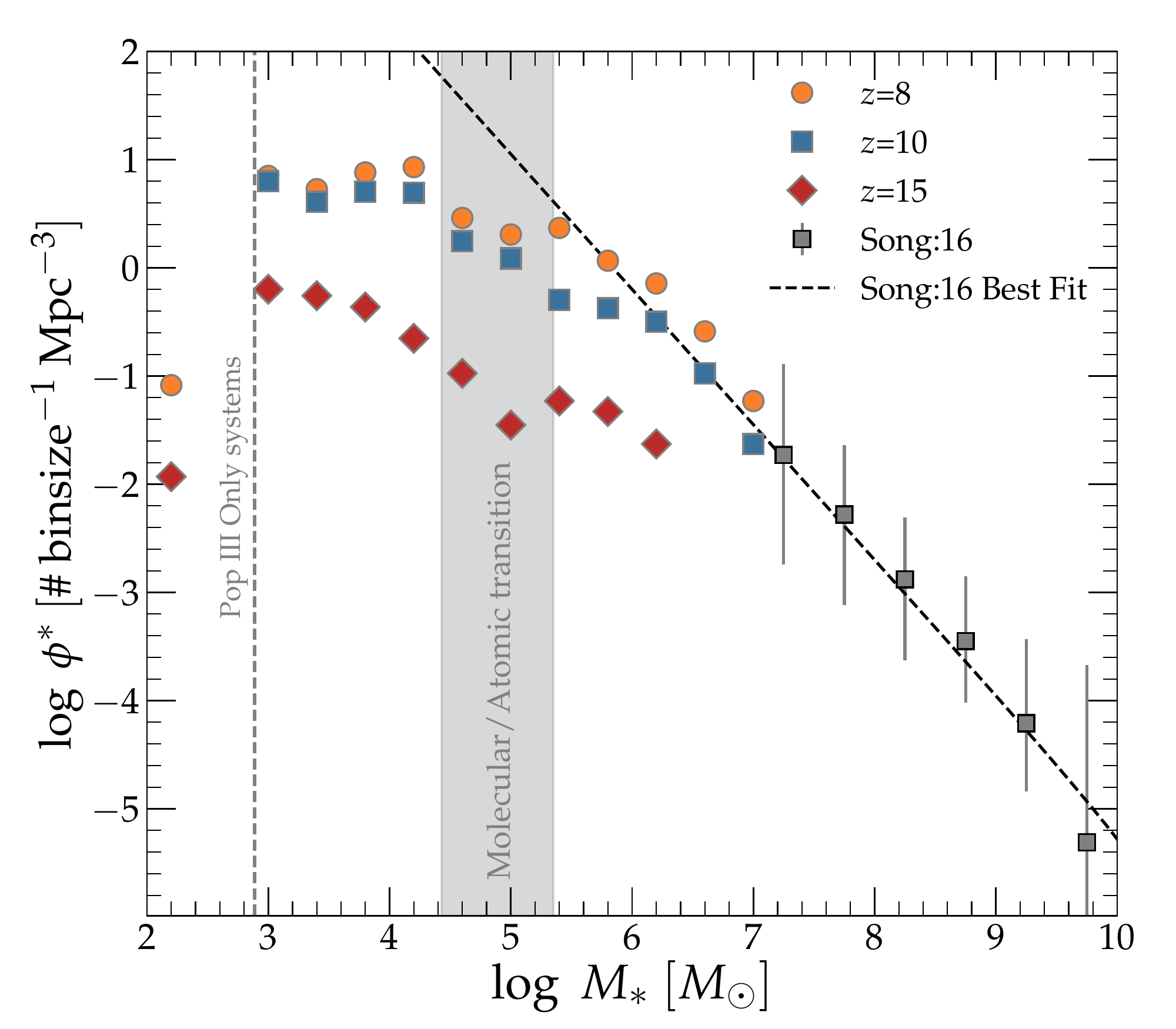}
\caption{Simulated galaxy stellar mass functions (SMF), shown for $z=8, 10, 15$ (orange circles, blue squares and red diamonds, respectively). The 
gray squares represent the observed SMF from \citet{Song:16}. We find excellent agreement with observations at $M_{*}\gtrsim10^5\ \Msun$.  However, for lower stellar masses we find a deviation (flattening) from the observed slope. This flattening coincides with the transition from molecular to atomic cooling haloes, at dark matter halo masses of $M_{\rm atomic}\approx 10^8\ ((1+z)/10)^{-3/2}\ \Msun$ (vertical gray shaded region).} 
\label{fig:smf}
\end{center}
\end{figure}

Further cross-checks on our star formation routines can be obtained by comparing our simulated galaxy stellar mass function (SMF), the number density of star forming objects within a given stellar mass bin, to observations. In Figure~\ref{fig:smf}, we compare our $z$=8, 10, 15 SMFs to the observed one at $z$=8, found in \citet{Song:16}. We construct the SMF by grouping star particles into galaxies, using a simple FOF algorithm and evolving the mass of each stellar population contained within, in accordance with its age and IMF. We find excellent agreement with both the normalization and  extrapolated slope of the observed SMF for $M_*<10^7\ \Msun$. On the low-mass end, at $M_*<10^5\ \Msun$, our simulated SMF is flattening and deviates from the empirical extrapolation, shown by the dashed black line. Conspicuously, the flattening seen in our simulated SMF occurs at a galaxy stellar mass, which corresponds to a dark matter halo mass of $M_{\rm halo}\sim 10^8\ \Msun$ at $z$=8. This is the mass scale where the transition between molecular dominated to atomic dominated cooling occurs (marked by the gray shaded region in Fig.~\ref{fig:smf}). It should be noted that the deviation from a power-law slope occurs in galaxies which contain $\gtrsim100$ star particles and reside in well resolved dark matter haloes. Therefore, we do not believe that the flattening feature is the result of limited resolution.  We discuss this interesting feature further in Section~\ref{subsec:turn} below.   

The simulated SMF exhibits significant evolution with increasing redshift, as the normalization decreases (orange circles compared to red diamonds). Because we do not capture the full range of masses in our meso-scale simulation box, it is unclear whether the SMF shape also evolves. In future work, we intend to increase the volume of our simulation in order to reproduce larger-mass systems, at which time we will be able to better quantify the evolution of both the normalization and the low-mass slope.
It should be noted that here, all galaxy stellar masses are calculated as the sum of Pop~II and Pop~III stars, taking into account the age and IMF of each star particle. For simplicity, the mass of stars which have evolved off the main sequence is no longer considered in the total mass of either component.

\subsubsection{UV Luminosity function}
\label{subsubsec:uvlf}
The final validation of our star formation model comes from the observed UV luminosity function (UVLF; the number of star forming objects within a given absolute magnitude bin per unit volume). While the SMF is straight forward to obtain from our simulations, it is more difficult to ascertain observationally.  Conversely, the UVLF is closer to a direct observable, though it requires additional steps and assumptions to derive from simulations.  We here discuss these assumptions, and compare our simulated UVLF to observations.

Producing a spectrum for a simulated star particle requires models for simple stellar populations (SSPs), over a range of stellar metallicities and age. For our Pop~II SSP, we adopt the \citet{Schaerer:02} low metallicity models, which have total stellar metallicities of $\log Z=-7.0, -5.0, -3.4, -3.0, -2.4, -2.1, -1.8$, ages in the range $10^4$ to $5\times10^7$~yr, and assume constant star formation histories.  For each star particle, we interpolate between the nearest two metallicities and ages. The total spectrum for each galaxy is then taken to be the sum of the individual spectra from each contributing star particle. To estimate the nebular emission, we employ a fixed value for the escape fraction of $f_{\rm esc}=0.10$ for all galaxies. This approximation is justified, given how incomplete our understanding of this key quantity still is. 

A slightly different approach is taken for the construction of our Pop~III spectra. To represent an individual Pop~III star we assume a simple blackbody curve, which has been shown to be a good approximation for a primordial star in \citet{Bromm:01c}. Our P3L star formation routine gives us a unique, randomly drawn stellar population for which each individual component mass is known. For simplicity we break down each Pop~III star particle into  four mass categories: PISN, High, Mid, Low. Each category is then assigned a temperature, radius and lifetime for which the blackbody spectrum and corresponding stellar luminosity is calculated. In Table~\ref{tbl:p3star}, we summarize the values assumed for those physical properties in each mass category. Similar to the Pop~II procedure, the total Pop~III spectrum is composed of each contributing star. These spectra are then added to the Pop~II  component, thus synthesizing the combined spectrum for a given galaxy.
\begin{table}
\caption{Physical properties of high mass Pop~III stars used to produce the blackbody spectra in our Pop~III SSP \citep{Schaerer:02}.}
\begin{center}
\begingroup
\setlength{\tabcolsep}{6pt} 
\renewcommand{\arraystretch}{1.2} 
\begin{tabular}{ccccc}
\hline
Type & Mass & T & Radius & Lifetime\\ 
	&[$\Msun$]& [K] & [R$_\odot$] & [Myr]	   \\
\hline
PISN & 145 & 95720	 &	4.80 & 1 \\
High & 90 & 93860	 &	3.91 & 3 \\
Mid & 24 & 70800	 &	1.85 & 10 \\
Low & 6 & 35000	 	 &	1.10 & 50 \\
\hline
\end{tabular}
\end{center}
\label{tbl:p3star}
\endgroup
\end{table}
Galaxy totals as well as the individual component spectra are then processed through a generic, top-hat filter, centered at $1500\ \angstrom$ with a total width of $100\ \angstrom$ to calculate the absolute UV magnitude.  We here neglect any dust extinction, such that $E(B-V)=0$, as both observations \citep{McLure:11,Dunlop:12,Finkelstein:12c,Bouwens:14a} and simulations \citep[e.g.][]{wilkins:16,Barrow:17,Zackrisson:17,Jaacks:17b} suggest that, on average, low-mass galaxies at high-$z$ contain insufficient dust to significantly redden their spectra.

\begin{figure*}
\begin{center}
\includegraphics[scale=0.47] {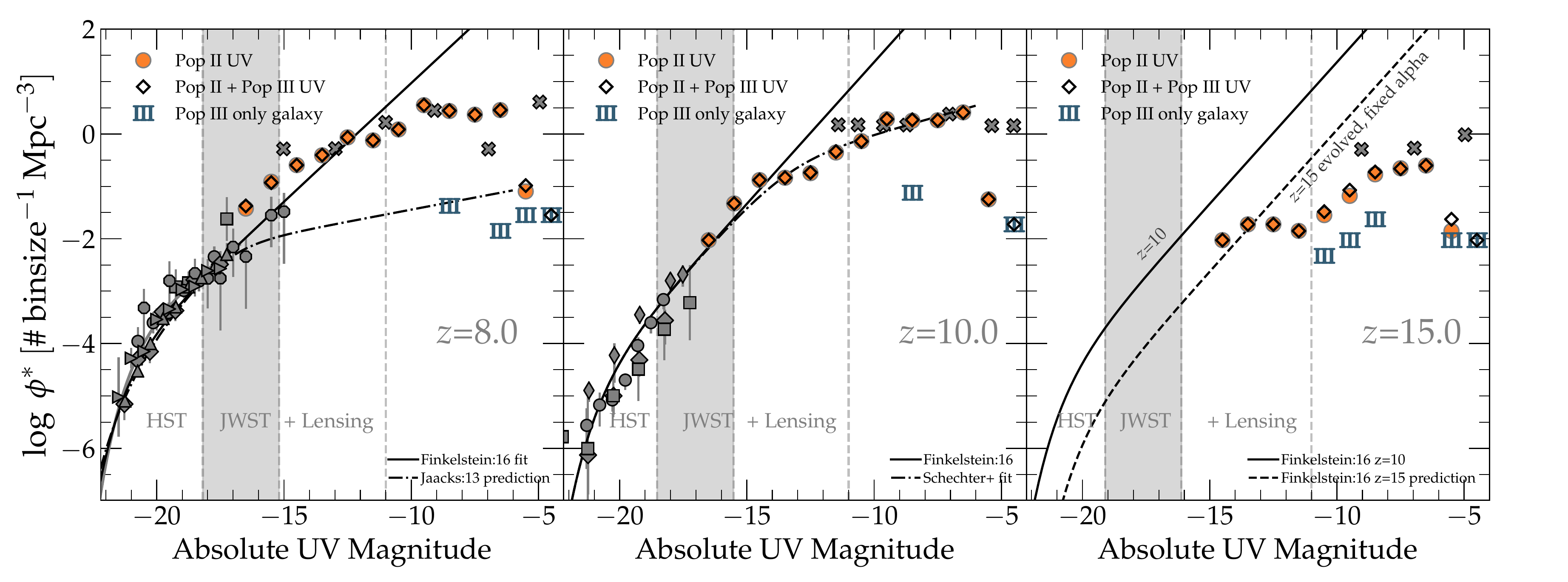}
\caption{UV luminosity functions (UVLFs) for $z$=8, 10, 15 (left, center, right, respectively). In each panel, we show the UV luminosity produced by Pop~II stars (orange circles) along with the UV luminosity produced by Pop~II + Pop~III stars (black diamonds). Sources with exclusively Pop~III-generated UV luminosity are shown by the blue ``III'' symbols. The solid black line in each panel is the Schechter function fit \citep{Schechter:76} from the 'reference' luminosity function found in \citet{Finkelstein:16}. We show excellent agreement with a wide range of observations at $z$=8 \citep[e.g.][]{Bouwens.etal:11a,Bouwens:14a,McLure:13,Schenker:13,Schmidt:14,Atek:15,Finkelstein:15}, and at $z$=10 \citep{Oesch:13,Oesch:18,Bouwens:15a,McLeod:16}.  Currently, there are no direct observations for $z$=15. However we include a prediction for the $z$=15 UVLF by evolving the Schechter fit parameters ($M^*$, $\phi^*$), according to \citet{Finkelstein:16} with a constant faint-end slope of $\alpha_{\rm UV}=-2.35$. We also include numerical studies from \citet{Wise:14} (gray 'x') and \citet{Jaacks:13} (dash-dotted black line). Throughout, the absolute UV magnitude was calculated with $E(B-V)=0.0$. In each panel, we show the approximate absolute UV magnitude limits for {\it HST} and {\it JWST}, assuming limiting AB magnitudes of $m_{\rm AB,lim}\approx 29$ and $32$, respectively. 
} 
\label{fig:uvlf}
\end{center}
\end{figure*}
 
In Figure~\ref{fig:uvlf}, we present the results of the above procedure in the form of the UVLF at $z=8, 10, 15$ (left, center, right).  In each panel, we show the UV luminosity produced by Pop~II stars (orange circles), along with that produced by Pop~II + Pop~III stars (open black diamonds).  Sources with exclusively Pop~III UV luminosity are marked by the blue ``III'' symbols. The solid black line in each panel is the Schechter function fit \citep{Schechter:76}, for the 'reference' luminosity function from \citet{Finkelstein:16}.
At $z=8, 10$, we show good agreement with the faint end ($M_{\rm UV}\sim-17$) of the observationally inferred UVLF (see Fig.~\ref{fig:uvlf} for references to select observations), and excellent agreement with numerical work from \citet{Wise:14} at the extreme faint end ($M_{\rm UV}>-15$).  Currently, there is no observational estimate for the $z=15$ UVLF. Therefore, we reproduce the Schechter fit to the $z=10$ observations to illustrate the relative evolution.  We also include a prediction for the $z$=15 UVLF by evolving the Schechter fit parameters ($M^*$, $\phi^*$) according to \citet{Finkelstein:16}, with a fixed faint-end slope of $\alpha_{\rm UV}=-2.35$.  While the agreement with our simulation results is encouraging, there remains much uncertainty regarding the evolution of the Schechter fit parameters.  Future direct observations with  {\it JWST} will allow for much better constraints at $z\gtrsim 10$.

An important aspect of this work is that we differentiate between the Pop~II and Pop~III contributions to the total galaxy spectra. This enables us to assess the separate contributions to the total galaxy luminosity. In each panel of Figure~\ref{fig:uvlf}, we can see that Pop~III only makes a minor contribution to the total UV luminosity (compare the black diamonds to the colored symbols), as the UVLF remains largely unchanged from the Pop~II-only case. This finding resonates with the SFRD results (see Sec.~\ref{subsubsec:sfrd}), where Pop~II dominates by more than a factor of 10 over Pop~III at these redshifts. We further quantify the relative contributions in Section~\ref{subsec:trans} below.  

We note a significant drop in the number density of galaxies with $M_{\rm UV}\gtrsim -10$, which approximately corresponds to $M_*\approx 10^4\ \Msun$ and $M_{\rm halo}\approx 10^6\ \Msun$. This deficit is in contrast to results found in \cite{Wise:14}, who use adaptive mesh refinement \citep[AMR;][]{Bryan:14} simulations to study highly-resolved first galaxies, finding flat number densities down to lower magnitudes. This contrast is possibly the result of our dark matter halo mass resolution limit of $M_{\rm halo}\approx 10^6\ \Msun$.  However, our Pop~III/II SSP masses ($M_{\rm *,III}\approx 500\ \Msun$, $M_{\rm *,II}\approx 1000\ \Msun$) are quite representative of typical, single star forming regions. Therefore, we can predict the magnitude where the UVLF is physically truncated, by calculating the UV magnitude for a single PopIII/II SSP, resulting in $M_{\rm UV, min,III}\sim-4.5$ and $M_{\rm UV, min,II}\sim-5.0$. Note that our Pop~III SSP is made up of randomly drawn components. Therefore, it is possible that a given SSP could reach even lower luminosities in rare cases. We are also assuming a zero-age SSP for both Pop~III and Pop~II for the purpose of this idealized calculation.

Recently, there has been much discussion in the literature regarding a possible turnover, or flattening, in the UVLF \citep[e.g.][]{Jaacks:13,Wise:14,Oshea:15,Livermore:17}. In the left panel of Figure~\ref{fig:uvlf}, we include the prediction from \citet{Jaacks:13}, who adopt a broken power-law functional form for the faint-end of the UVLF. In the equivalent luminosity form, this can be written as
\begin{equation}
\Phi(L)=\phi^*\left(\frac{L}{L^*}\right)^{\alpha_{\rm UV}} \exp \left(-\frac{L}{L^*}\right)\left[1+\left(\frac{L}{L_{\rm turn}}\right)^{\beta_{\rm UV}}\right]^{-1}.
\label{eq:lf+}
\end{equation}
Here, ${L_{\rm turn}}$ constrains the turnover luminosity, and $\beta_{\rm UV}$ the subsequent flattening.  When converted to its UV magnitude formulation, one has
\begin{multline}
\Phi(M) = 0.4\ln 10 \phi^*  10^{0.4(M_{\rm UV}-M)(1+\alpha_{\rm UV})}\exp\left(-10^{0.4(M_{\rm UV}-M)}\right)\\
\times \left[1+10^{(M_{\rm UV, turn}-M)\beta_{\rm UV}} \right]^{-1}\mbox{\ .}
\label{eq:lfmag+}
\end{multline}
However, based on the recent observation by \citet{Livermore:17}, it appears that the \citet{Jaacks:13} prediction fails to properly represent either the observed data, or the simulation results presented here (see dash-dotted line in $z$=8 panel).  This is likely due to a resolution in the earlier study which was insufficient to accurately model star formation in low mass molecular cooling haloes. In the center panel of Figure~\ref{fig:uvlf}, we provide a fit to the combined data set (i.e. observations and this work), using Equation~\ref{eq:lfmag+}.  We find that when we fix the standard Schechter parameters to the $z$=10 values found in \citet[$\log \phi^*=-4.13,\ M_{\rm UV}=-20.25,\ \alpha_{\rm UV}=-2.35$]{Finkelstein:16}, the turnover seen in our simulated data can be constrained by $M_{\rm UV, turn}=-13.4\pm1.1$ and $\beta_{\rm UV}=-1.0\pm0.2$.
We provide an extended discussion on this topic in Section~\ref{subsec:turn}.

\subsection{Cosmic enrichment evolution}
\label{subsec:evo}
Above, we have explored the global properties of the galaxies and haloes found in our simulation volume. We now focus on how these properties evolve over cosmic time. In particular, we will investigate the transition from a Pop~III dominated Universe to one dominated by Pop~II star formation. By investigating this transition, we lay the foundation for discussions regarding the overall legacy of Pop~III star formation in the cosmological context, and the probability of observing the signature of the first stars (see Sec.~\ref{sec:predict}).

\begin{figure}
\begin{center}
\includegraphics[scale=0.40] {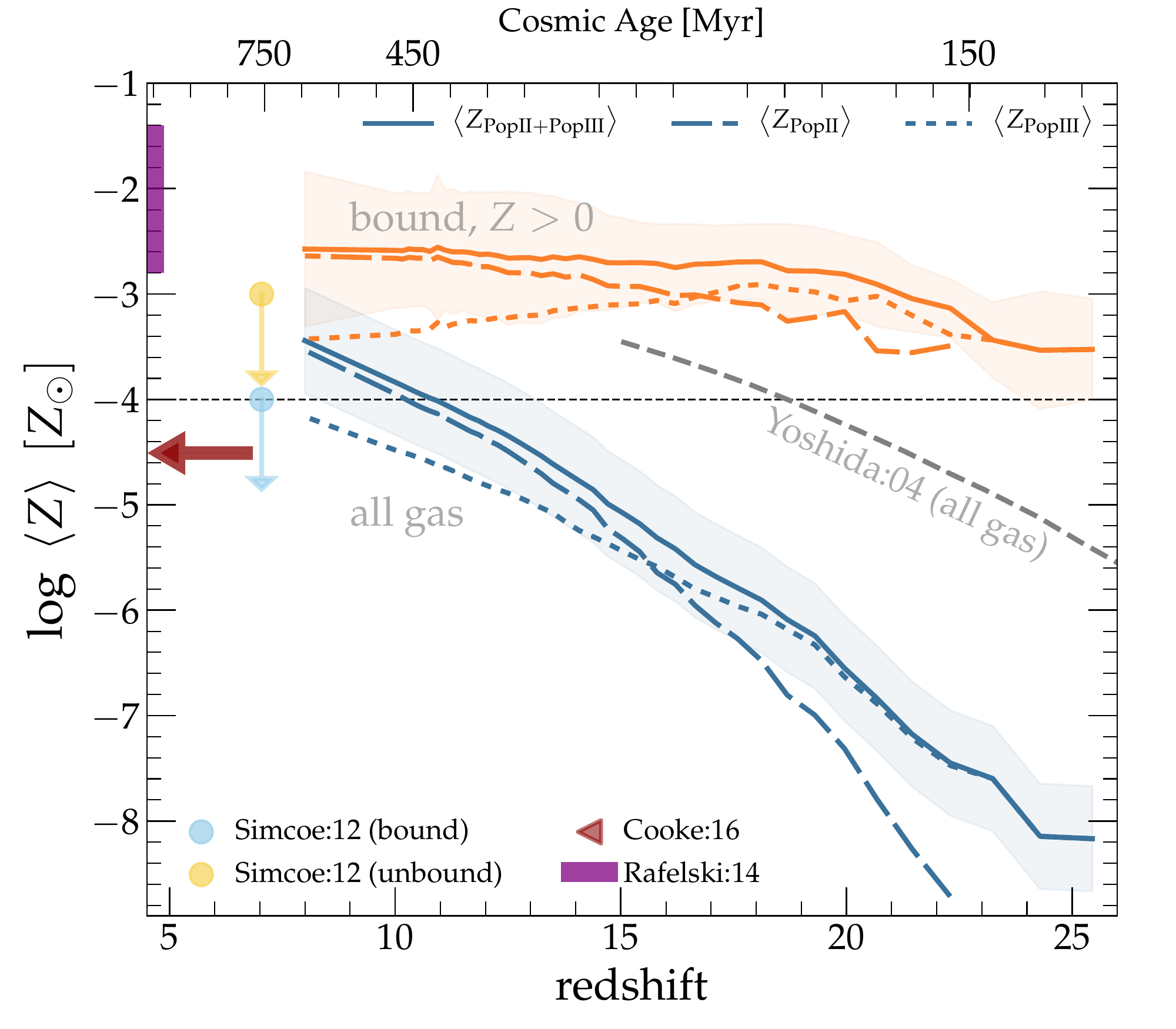}
\caption{Mean metallicity evolution calculated for enriched, bound star forming regions (orange) and the entire simulation volume (blue).  The dashed and dotted lines represent metals produced by Pop~II and Pop~III events, respectively, and the solid lines denote the total metallicity (i.e. Pop~II + Pop~III). We also present the predicted metallicity evolution from \citet{Yoshida:04}, where {\it all} Pop~III stars were considered to end their lives as PISN, thus imparting a maximum amount of metals back into their environment. Observations of high redshift DLA systems are indicated by the yellow/blue circles \citep[$z=7.04$;][]{Simcoe:12}, and the purple range \citep[$z>4.7$;][]{Rafelski:14}. The yellow circle is the estimate based on the assumption of an unbound medium, whereas the blue point is for a bound structure.  Both are considered upper limits.  The red arrow points to a lower redshift observation by \citet{Cooke:16}.} 
\label{fig:metals}
\end{center}
\end{figure}

\subsubsection{Mean metallicity}
\label{subsubsec:metal_evo}
In Figure~\ref{fig:metals}, we present the mean metallicity evolution for both enriched, bound systems (dark matter haloes) and all gas in our simulation volume.  As seen in our previous work which focused only on enrichment via Pop~III star formation \citep{Jaacks:17a}, bound systems (solid orange line) immediately jump above the critical metallicity line ($Z_{\rm crit}=10^{-4}\ \Zsun$) at the onset of star formation ($z\sim26$). The total metallicity ($Z_{\rm Pop~II}+Z_{\rm Pop~III}$) for bound systems then rises slowly to a plateau value of $\log \left<Z_{\rm bound,total}/\Zsun\right>\approx-2.5$.  This plateau suggests that a near-equilibrium between accretion of metal free gas and ongoing star formation has been established early on.

The mean total metallicity for all gas particles in our simulation volume (solid blue line) rises from an initial value of $\log \left<Z_{\rm all,total}/\Zsun\right>\approx -8$ at $z\sim26$ to a value of $\log \left<Z_{\rm all,total}/\Zsun\right>\approx -3.5$ at $z=8$.  The mean metallicity in our simulation volume does not cross $Z_{\rm crit}$ until $z\sim11$.  This corresponds to the slight `flattening', seen in the Pop~III SFRD at $z\lesssim13$ as the available reservoir of high-density, Pop~III star forming gas is depleted. It is interesting to note that, while depleted, there remain substantial pockets of low-metallicity or metal-free gas to sustain ongoing Pop~III star formation over the entire redshift range studied here.  This suggests that, globally, Pop~III star formation is not terminated by solely metal enrichment processes, in contrast to earlier predictions, such as in \citet{Yoshida:04}.

\subsubsection{Metal volume filling fraction}
\label{subsubsec:vff}
\begin{figure}
\begin{center}
\includegraphics[scale=0.40] {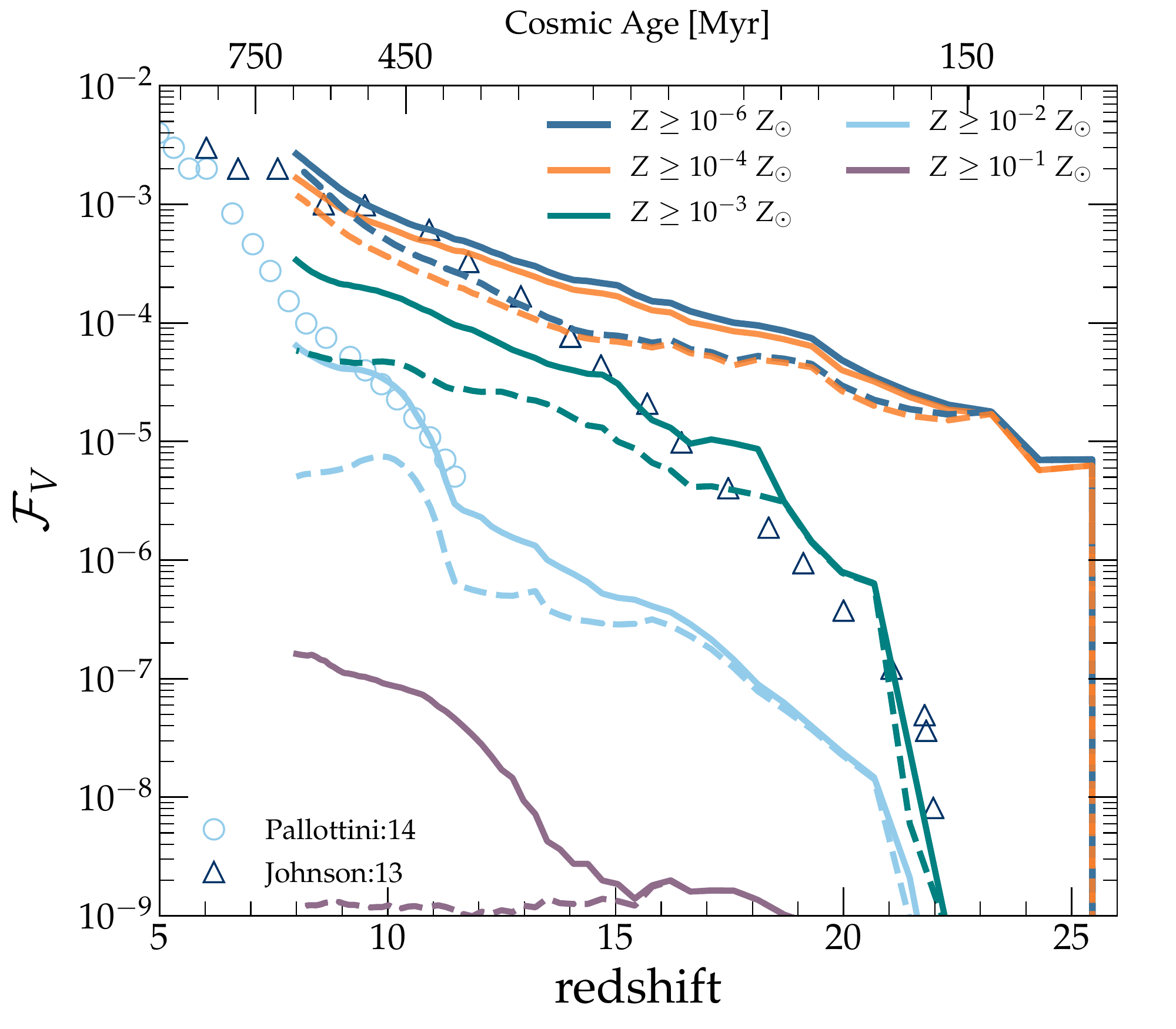}
\caption{Metal volume filling fraction for both Pop~II (solid lines), and Pop~III (dashed lines) produced metals, shown for metallicity thresholds varying between $10^{-6}\leq Z/\Zsun \leq 10^{-1}$. We also reproduce results from \citet{Pallottini:15} via the cyan open circles, which can be directly compared to the solid cyan line from our work.  The blue open triangles shows results from \citet{Johnson:13} for the volume filling fraction of all metals.} 
\label{fig:vff}
\end{center}
\end{figure}

To better quantify the spatial extent of both Pop~II and Pop~III metal enrichment, we in Figure~\ref{fig:vff} present volume filling fractions, $\mathcal{F}_V$, for a range of metallicity thresholds ($10^{-6}\leq Z/\Zsun \leq 10^{-1}$), as a function of redshift. As expected, we see a trend of increasing metallicity for both Pop~III (dashed lines) and Pop~II (solid lines) produced metals, with Pop~II enrichment delayed by comparison.  We also discern the same trend as in Figure~\ref{fig:metals} with Pop~II enrichment catching up and surpassing Pop~III between $z\approx15-20$, depending on the threshold value. Evidently, the cosmic buildup of metallicity with higher thresholds is delayed until later redshifts. For example, there is no volume element with $Z>10^{-1}\ \Zsun$ until $z\approx 19$, whereas regions of the simulation are enriched beyond $Z>10^{-6}\ \Zsun$ immediately upon the onset of star formation at $z\approx 26$ . 
By the end of our simulation, at $z$=7.5, we find $\mathcal{F}_V\approx10^{-3}$ for $Z\geq10^{-6}\ \Zsun$.  This compares very favorably to results from \cite{Johnson:13}, who find a similar value for all enrichment in their simulation volume.  We deviate from these results at $z\gtrsim15$ due to our higher SFRD at these redshifts. We find that only $\sim 10^{-4}$ of the simulation volume is enriched to beyond the critical metallicity for the Pop~III to Pop~II transition, $Z_{\rm crit}=10^{-4}\ \Zsun$ by $z=7.5$, suggesting that there remains a large fraction of gas which has not been enriched by either population.  This gas is potential fuel for ongoing Pop~III star formation events.  Our results are also consistent with \citet{Pallottini:15} for a threshold value of $Z\geq10^{-2}\ \Zsun$.

\subsection{Pop~III/II transition}
\label{subsec:trans}

\begin{figure}
\begin{center}
\includegraphics[scale=0.40] {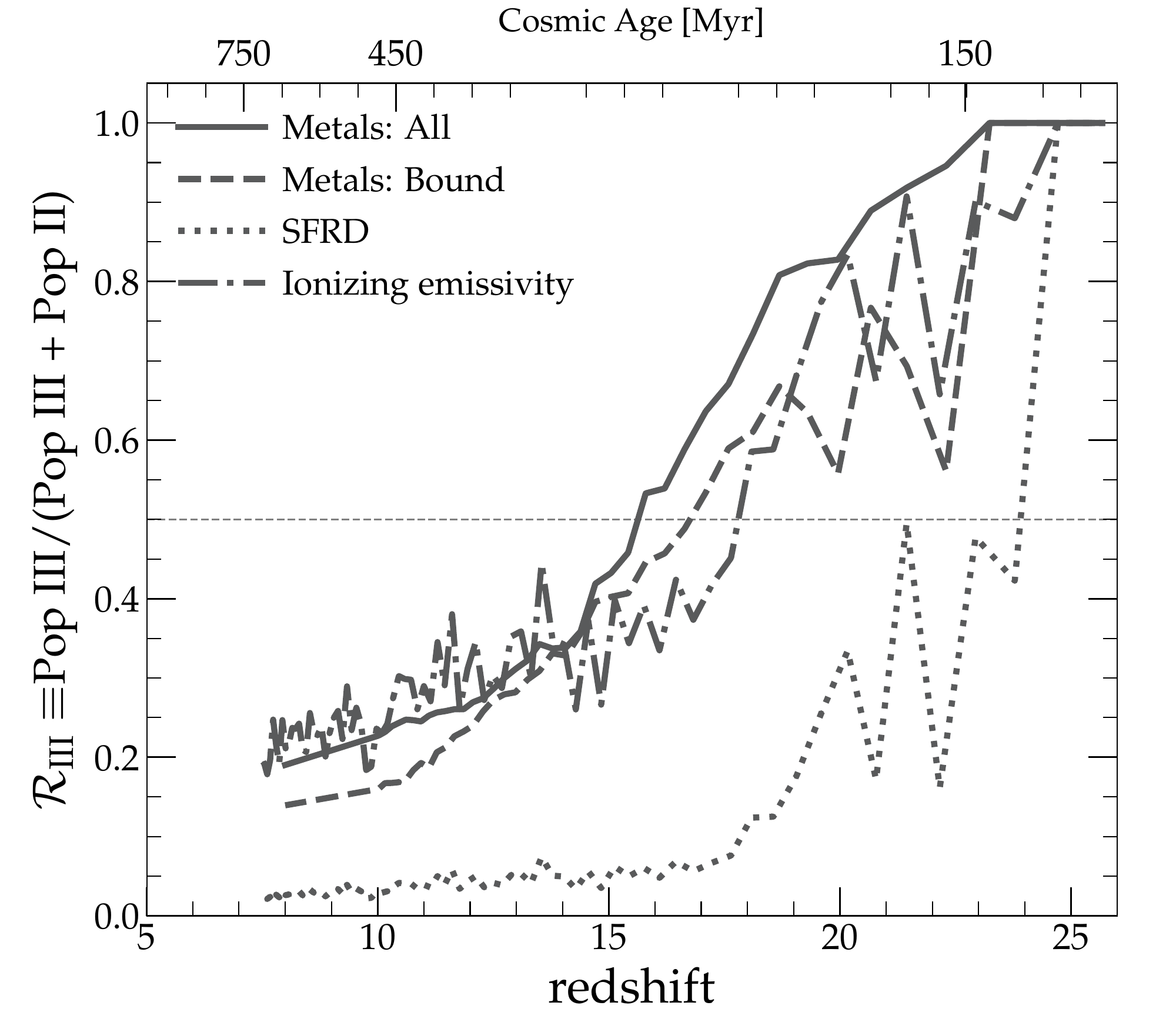}
\caption{Ratio ($\mathcal{R_{\rm III}}$) of metals, SFRD and ionizing emissivity, contributed by Pop~III to the total. The horizontal line represents the fiducial break-even level, $\mathcal{R_{\rm III}}=50\%$. In terms of metal enrichment, it is evident that in bound star-forming structures (i.e. haloes), Pop~III is overtaken by Pop~II earlier ($z\sim17$) than for gas in the diffuse IGM ($z\sim15$). We also see that Pop~II star formation quickly dominates over Pop~III ($z\sim20$). However, due to the more efficient ionizing photon production, Pop~III star formation still contributes $\mathcal{R_{\rm III}}=20\%$ to the total at $z\sim7.5$.} 
\label{fig:ratio}
\end{center}
\end{figure}

We now wish to understand in more detail the relative contributions of Pop~III and Pop~II to star formation and key feedback processes, and in particular the epoch when the latter begins to dominate. For this purpose, we define the ratio 
\begin{equation}
\mathcal{R}_{\rm III}\equiv \frac{\rm Pop~III}{\rm Pop~III+Pop~II}.
\end{equation}
In Figure~\ref{fig:ratio}, we show $\RIII$ for three different star formation tracers: the SFRD, metals (bound, all), and ionizing emissivity. To guide the eye, we add the horizontal line, denoting the $\RIII=0.50$ value.
From Figures~\ref{fig:metals} and \ref{fig:ratio}, it can be seen that Pop~III metal enrichment dominates at $z\gtrsim 17$ for bound systems (dashed line), and $z\gtrsim 15$ for all gas (solid line), after which Pop~II takes over.  By $z\approx8$, only $15\%$ (bound systems) and $20\%$ (all gas) of the metals originated from Pop~III sources.  We also see that bound systems cross the $\RIII=0.50$ line earlier than the remainder of the volume. This is the case, because bound systems will be enriched first, as the hosts for star formation.

In Figures~\ref{fig:sfrd} and \ref{fig:ratio} it is apparent that Pop~II star formation quickly follows the first burst of Pop~III star formation (within $\sim 15$ Myr). While the two star formation modes initially contribute at a similar level, by $z\sim 20$ Pop~II clearly dominates (dotted line). More specifically, by $z=7.5$, Pop~III contributes $\sim5\%$ to the total SFRD, which is consistent with numerical results from \citet{Pallottini:15}, who find a ratio of $8\%$ at $z\sim 7$. However, since Pop~III stars are, on average, more massive and thus hotter, they produce an order of magnitude more ionizing photons per stellar baryon \citep{Bromm:01c}. This is reflected in Pop~III stars still contributing $\sim 20\%$ of the ionizing photon budget at $z=7.5$, which is overall consistent with previous studies, finding contributions of $\sim10\%$ \citep[e.g.][]{Ricotti:04,Greif:06,Wise:12,Paardekooper:13}. Again, our slightly higher value can be attributed to the slightly larger Pop~III SFRD, predicted here.

\subsection{Where does Pop~III occur?}
\label{subsec:occur}

\begin{figure*}
\begin{center}
\includegraphics[scale=0.40] {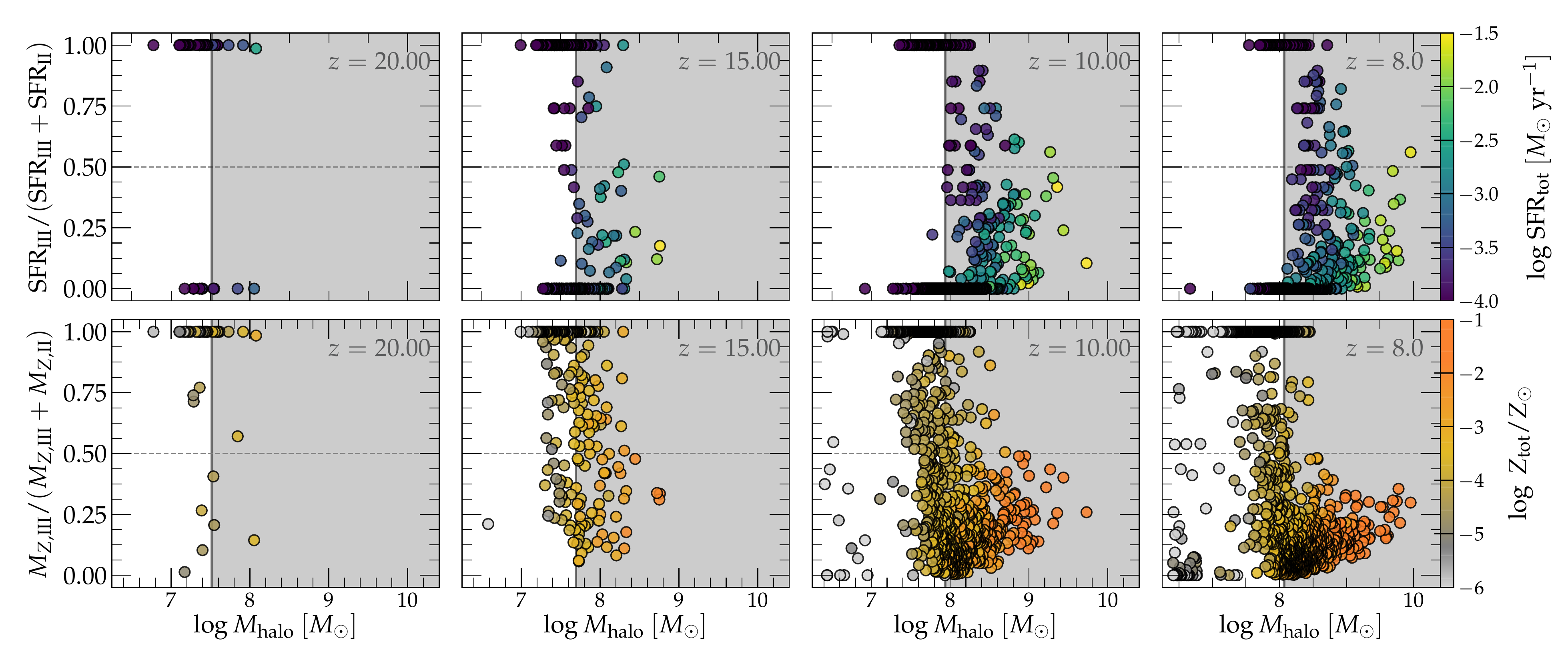}
\caption{ {\it Top row:} Ratio of Pop~III SFR to total SFR (Pop~II+Pop~III), as a function of dark matter halo mass for $z=20, 15, 10, 8$.  The colors in each panel represents the total SFR for each dark matter halo.  Pop~II/III SFRs are determined by summing the stellar mass which has formed within the previous 50/10~Myr. The shorter Pop~III time-average reflects the shorter lifetime of Pop~III SSPs.  {\it Bottom row:}  Ratio of Pop~III metal mass ($M_{Z,\rm III}$) to total metal mass ($M_{Z,\rm III} + M_{Z,\rm II}$), as a function of dark matter halo mass for the same redshifts as above. Now, the colors in each panel represent total metallicity ($Z_{\rm tot}/\Zsun$) in solar units for each dark matter halo.  The gray shaded area in each panel represents the halo masses which are massive enough to enable atomic hydrogen cooling.} 
\label{fig:mh_sfr}
\end{center}
\end{figure*}

Above, we have demonstrated that Pop~III star formation, on average, is quickly dominated by Pop~II (see Figs.~\ref{fig:sfrd} and \ref{fig:ratio}).  However, it is clear from Figure~\ref{fig:sfrd} that Pop~III star formation has not been completely terminated by $z\sim7.5$. This leads to the question: Is the ongoing Pop~III star formation occurring in isolated, primordial haloes or pristine regions of Pop~II dominated haloes? To answer this question, we examine the ratio of star formation rates (SFRs), defined as
\begin{equation}
\mathcal{R_{\rm SFR,III}}\equiv\frac{\rm SFR_{III}}{\rm SFR_{III}+SFR_{II}}.
\end{equation}

In the top row of Figure~\ref{fig:mh_sfr}, we present the $\mathcal{R_{\rm SFR,III}}$ as a function of dark matter halo mass at $z=20, 15, 10, 7.5$.  We calculate the SFR by including stars formed within the past $50$~Myr for Pop~II, and $10$~Myr for Pop~III. Furthermore, the color bar represents the total SFR for each halo. The intuition here is that haloes which are experiencing Pop~III-only star formation have $\mathcal{R_{\rm SFR,III}}=1.0$, whereas haloes which are experiencing solely Pop~II star formation will have $\mathcal{R_{\rm SFR,III}}=0.0$.  
At early times, $z\sim 20$, low-mass haloes, $M_{\rm halo}\lesssim10^{7.5}\ \Msun$, are Pop~III dominated, with only a few systems experiencing Pop~II star formation. With increasing age of the Universe (left to right), we record a larger number of systems which become Pop~II dominated. The Pop~III/II transition is highly correlated with the transition in mass between molecular cooling and atomic cooling haloes (indicated by the gray shaded area). This in turn is consistent with the current star-formation paradigm, where Pop~III stars form in low-mass minihaloes, thus planting the seed for subsequent Pop~II stars.

Interestingly, the highest mass haloes in each panel exhibit $\mathcal{R_{\rm SFR,III}}<1.0$, which indicates that both modes of star formation are ongoing. This is likely the result of pristine, neutral gas being accreted to within the virial radius of a halo that already contains Pop~II stars, without experiencing significant mixing and shock heating.  The survival of any pockets of in-falling primordial gas depends on the detailed physics of turbulence-driven mixing of heavy elements \citep[e.g.][]{SmithB:15,Sluder:16,MJ:17, Sarmento:18}.  Those fine-grained hydrodynamical mixing processes are not resolved here, such that we may overestimate the occurrence of Pop~III star formation inside the more massive host haloes. However, on the scale of a pre-stellar clump, which is resolved here, diffusion and mixing of metals may be too slow to penetrate deep enough into the clump to prevent Pop~III star formation there \citep[e.g.][]{Cen:08}. Clearly, this needs to be addressed further with future higher-resolution simulations.  

\begin{figure}
\begin{center}
\includegraphics[scale=0.40] {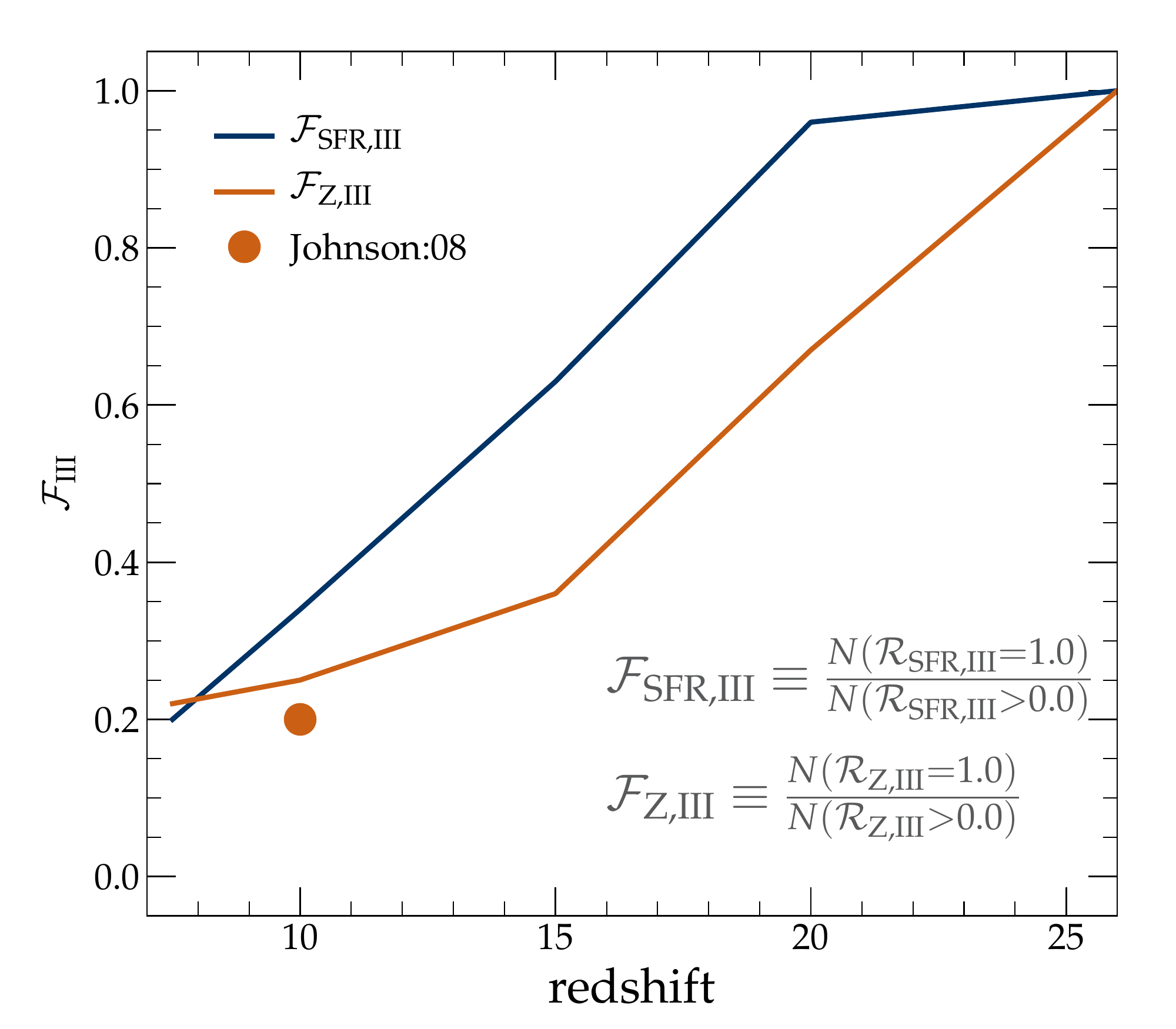}
\caption{Fraction of Pop~III-only systems. We show the fraction of Pop~III star forming haloes which have yet to experience Pop~II star formation, as a function of redshift (blue line). This is effectively the fraction of isolated haloes hosting Pop~III, as opposed to mixed-population ones. Similarly, we show the fraction of haloes which only contain Pop~III-generated metals, compared to all enriched systems (orange line). Both curves are extracted from the data presented in Fig.~\ref{fig:mh_sfr}. We include an estimate at $z\sim 10$, provided in the numerical work by \citet{Johnson:08} via the orange circle.
} 
\label{fig:p3_ratio}
\end{center}
\end{figure}

This so-called cold-mode accretion \citep{Birnboim:03,Keres:05,Dekel:06}  then provides the fuel for ongoing Pop~III star formation.  Thus, there is a prominent class of haloes with a mixed star formation mode. We do, however, wish to assess the fraction of haloes, with ${\rm SFR_{III}}>0.0$, which have experienced no Pop~II star formation, defined as
\begin{equation}
\mathcal{F}_{\rm III}\equiv\frac{N({\mathcal{R_{\rm SFR,III}}=1.0})}{N({\mathcal{R_{\rm SFR,III}}>0.0})}.
\label{eq:p3_frac}
\end{equation}
This quantity can be interpreted as the fraction of isolated haloes, hosting Pop~III-only, among all Pop~III-forming haloes. As can be seen in Figure~\ref{fig:p3_ratio}, there is a clear trend of lower $\mathcal{F}_{\rm III}$ with decreasing redshift. When our simulation ends at $z\simeq 7.5$,  $\sim20\%$ of Pop~III star formation is occurring in isolated, Pop~III-only, haloes, with ratios of 34\%, 63\%, and 96\% at $z=10, 15, 20$, respectively.

It is interesting to note that in the top panel of Figure~\ref{fig:mh_sfr}, we see Pop~II star formation occurring in dark matter haloes which are below the atomic cooling limit, indicating that these star forming regions are cooling via channels other than collisional excitation of atomic \ion{H}{I} and \ion{He}{I}. Cooling in haloes with $M_{\rm halo}<M_{\rm atomic}$ ($T_{\rm vir}<10^4$ K) must then be dominated by fine-structure transitions in metal enriched gas, in agreement with the results in \cite{Wise:14}.

The bottom row of Figure~\ref{fig:mh_sfr} shows the ratio of mass in Pop~III-generated metals to total metal-enriched mass, $\mathcal{R_{\rm Z,III}}\equiv{\rm M_{Z,III}/(M_{Z,III}+M_{Z,II})}$, as a function of halo mass, where the color bar indicates the total metallicity of each halo. Here, the trend is very similar to the SFR one, which is expected since metals are the direct result of star formation events. By inspecting the color bar, we can also discern a clear trend of increasing metallicity with increasing halo mass, with the highest mass halo, $M_{\rm halo}\approx10^{10}\ \Msun$, containing $Z_{\rm tot}\approx10^{-1}\ \Zsun$. We also notice that at or near the molecular/atomic cooling mass, haloes are enriched to beyond the critical Pop~III/II metallicity ($Z_{\rm crit}=10^{-4}\ \Zsun$).

In Figure~\ref{fig:p3_ratio}, the orange line corresponds to the fraction of haloes which contain only Pop~III-generated metals, compared to all enriched haloes. We calculate this quantity again with Equation~\ref{eq:p3_frac}, but now using metallicity instead of SFR. The offset between the blue and orange lines in Figure~\ref{fig:p3_ratio} is due to the fact that Pop~III metals persist, whereas the Pop~III SFR is temporary. Thus, a Pop~II dominated halo will always contain Pop~III metals, but may not experience ongoing Pop~III star formation. Therefore, the denominator of Equation~\ref{eq:p3_frac} will typically be larger in the case of metals, leading to a lower ratio.

\section{Frontier observations}
\label{sec:predict}

We find that, at each redshift studied, there are galaxies which exclusively consist of Pop~III stars (see blue ``III'' symbols in Fig.~\ref{fig:uvlf}).  However, detecting these `pure' Pop~III galaxies is beyond the capabilities of {\it JWST}, even when lensing is utilized, allowing us to reach $M_{\rm UV}\lesssim -11$). It is thus unlikely that upcoming surveys will be able to directly detect such Pop~III-only, or Pop~III-dominated, galaxies. There are, however, several empirical avenues to indirectly probe these systems, among them are absorption studies of the diffuse IGM, and searches for transient events at high-$z$, such as SNe and gamma-ray bursts (GRBs).

\subsection{Probing the IGM metallicity}
The search for systems which have been enriched only with Pop~III metals has intensified in recent years. Prime examples are the vigorously debated luminous CR7 Lyman-$\alpha$ source, which was initially thought to exhibit ultra-low metallicity \citep{Sobral:15}, and the \citet{Simcoe:12} damped Lyman-$\alpha$ (DLA) system. It is useful to utilize our simulations to examine the metal enrichment as a function of environment, to determine the observability of ultra-low or zero metallicity systems at $z\gtrsim 7$.

\begin{figure}
\begin{center}
\includegraphics[scale=0.43] {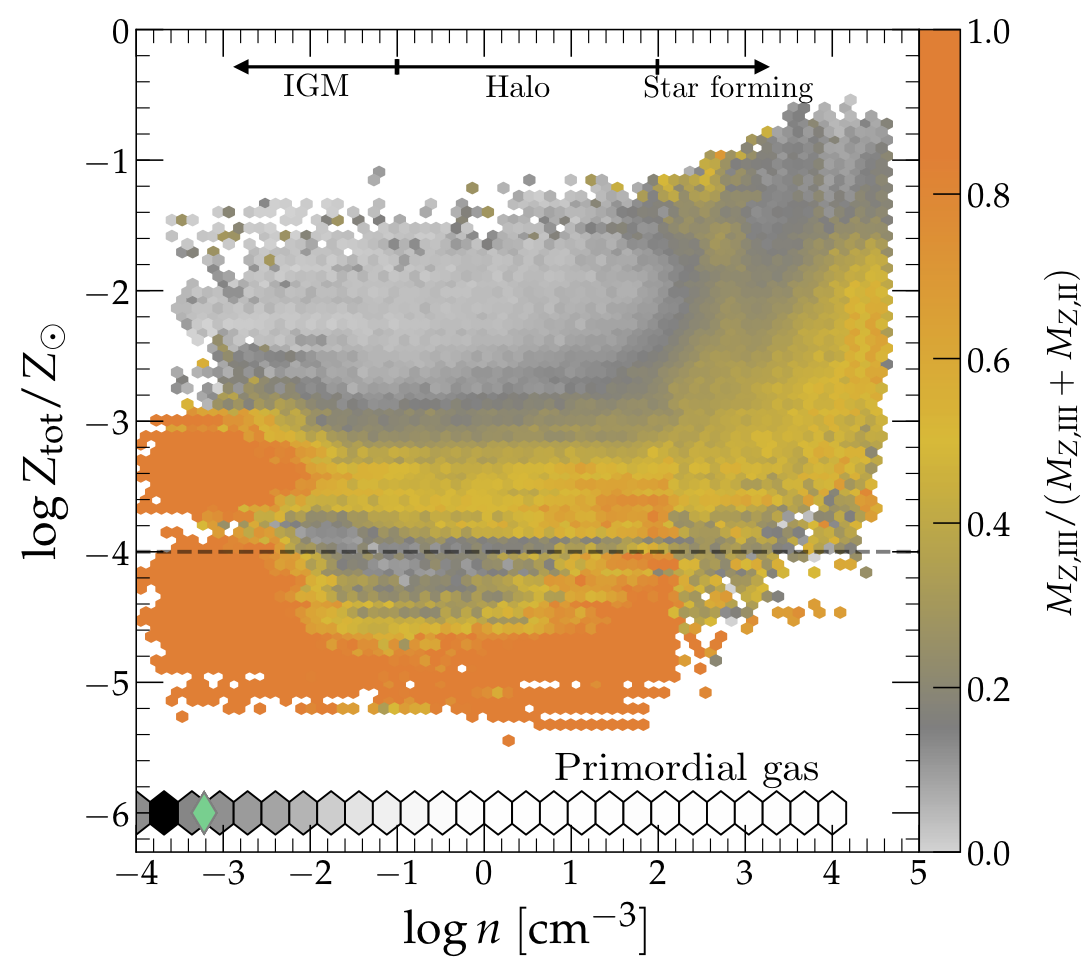}
\caption{Metallicity as a function of number density for each gas particle in our simulation volume.  The color bar represents the mass fraction of Pop~III metals contained in each particle.  The dashed horizontal line indicates the critical metallicity for the Pop~III to Pop~II transition ($Z_{\rm crit}=10^{-4}\ Z_\odot$).  The approximate delineations between regions within our volume (IGM, halo, and star forming) are indicated at the top of the figure.  In order to include gas which has yet to be enriched by either Pop~II or Pop~III metals we artificially place this gas at $\log Z/\Zsun=-6$, where the black to gray shading indicates particle frequency, with black being the highest.  The cyan diamond indicates the median number density of the primordial gas ($\log n\simeq -3.2\ \pccc$).
} 
\label{fig:ndens}
\end{center}
\end{figure}

In Figure~\ref{fig:ndens}, we show the total gas metallicity as a function of number density, for all gas particles in our simulation volume.  The color represents the ratio of Pop~III metals contained in each hexagonal pixel.  We also artificially place gas with zero metallicity at $\log Z/\Zsun=-6$, and indicate particle frequency with shades from black (highest) to white (lowest). As one would expect, the region above our star formation threshold of $n_{\rm th}=100\ \pccc$, and above $Z_{\rm crit}$ (dashed line), shows a mix of Pop~II + Pop~III metal enrichment, with higher metallicities being dominated by Pop~II metals. Conversely, below $Z_{\rm crit}$, Pop~III metals dominate.  It is also clear that Pop~III metals permeate throughout all of the regions indicated (IGM, halo, star forming). This renders identifying regions where observations could look for systems that are exclusively enriched by Pop~III particularly challenging, as Pop~III metals from these systems span the entire range presented here.

It is also useful to identify regions which have experienced zero enrichment (primordial gas). In Section~\ref{subsubsec:vff}, we found that only $\sim 10^{-3}$ of our simulation volume has been enriched by $z=7.5$.  As indicated by the black-gray shaded region in Figure~\ref{fig:ndens}, the vast majority of the primordial gas is contained in the low-density IGM  ($n<10^{-2}\ \pccc$), with a mean density of $\log \left<n_{\rm pri}\right>\simeq -3.2\ \pccc$ (cyan diamond in Fig.~\ref{fig:ndens}). While primordial gas exists over the entire dynamic range of our simulated volume, it will be difficult to detect when contained within dark matter haloes, as most lines-of-sight will also contain metal-enriched gas. Therefore, the best opportunity to detect primordial gas with absorption spectroscopy may be at the interface between the diffuse IGM and filamentary structures of the cosmic web. Future observations with the upcoming suite of extremely large, 30-40m class telescopes on the ground may be able to push existing limits on the Lyman-$\alpha$ forest into the regime of the chemically pristine IGM.    
Note that the number density of each particle is directly tied to the hydrodynamical smoothing length of the particle.  While {\small GIZMO} incorporates an adaptive hydrodynamical smoothing length, it imposes a maximum of $0.45$~kpc (comoving), which is why we have a buildup of gas at $n\approx10^{-4}\ \pccc$.

\subsection{Transient event rate}
Transient events, such as SNe and GRBs, may be our best mechanism to probe low-density gas at $z\gtrsim 10$, as they act as background flashlights to illuminate foreground systems \citep{Wang:12}. Therefore, estimating the production rate for these events may prove useful for future deep-field surveys. To first order, the rates for PISNe and the less-extreme core-collapse SNe (CCSNe) can be calculated by leveraging our Pop~II/III SFRDs, together with information about their respective IMFs, as \citep[e.g.][]{Hummel:12}
\begin{eqnarray}
\frac{dN}{dt_{\rm obs} dz d\Omega}&=&\zeta_{\rm IMF}\frac{dN_*}{dt_{\rm obs}dV}\frac{dV}{dzd\Omega}  \nonumber \\
								 &=&\zeta_{\rm IMF}\frac{1}{(1+z)}\frac{dN_*}{dt_{\rm em}dV}r^2\frac{dr}{dz}.	
\end{eqnarray}
Here, $dN_*/dt_{\rm em}dV$ is the star formation rate per comoving volume element (SFRD), $\zeta_{\rm IMF}$ accounts for the fraction of the IMF which falls within the appropriate mass range for each transient, and $r=r(z)$ is the comoving distance to redshift $z$. The result is the number of events per unit time per unit redshift per solid angle.

\begin{figure}
\begin{center}
\includegraphics[scale=0.40] {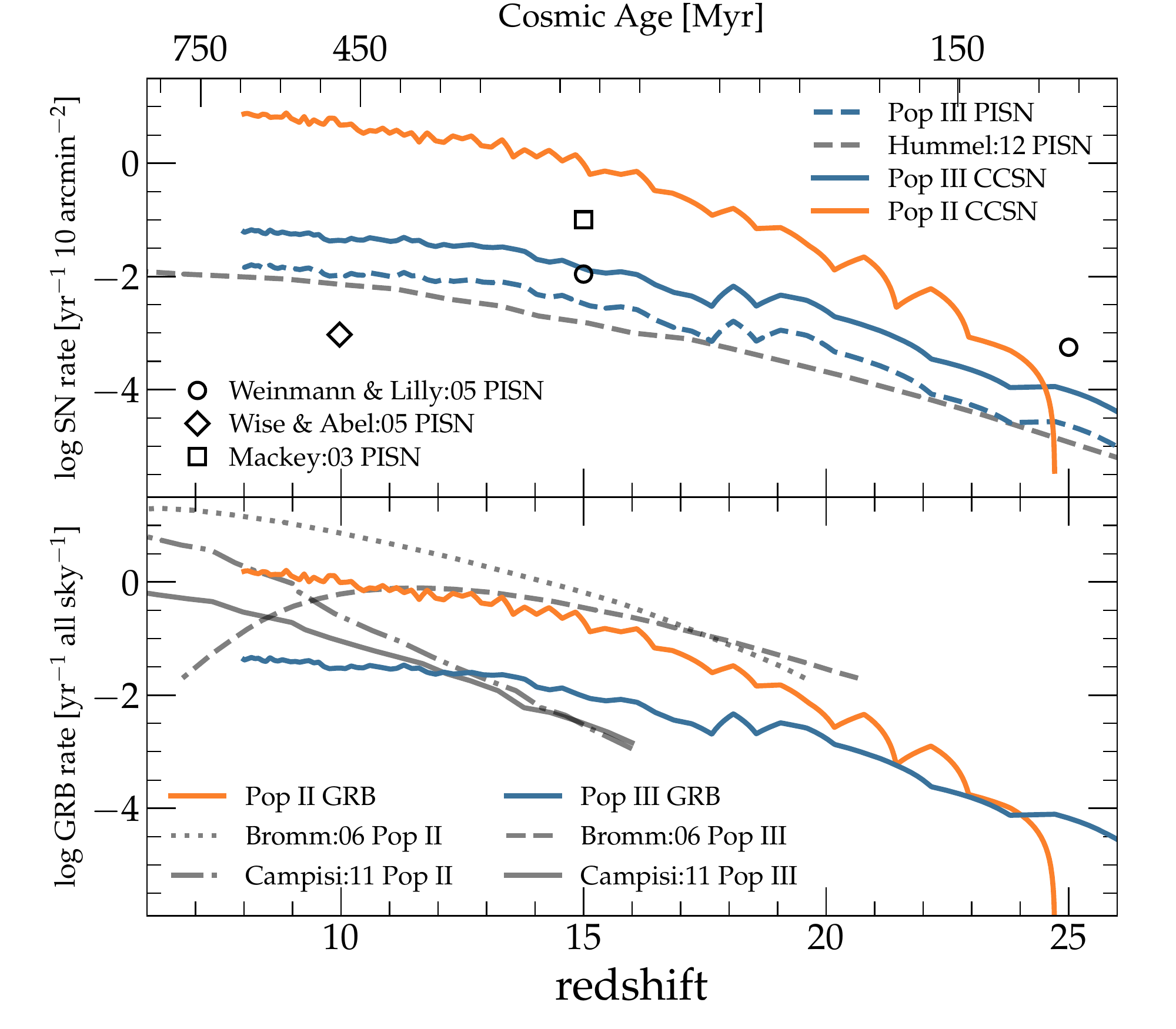}
\caption{Rate of transient events for both Pop~II (orange lines) and Pop~III (blue lines). {\it Top:} SN rates, both for CCSNe and PISNe. Note that we employ units of events per year per 10 arcmin$^2$, the approximate field of view of a {\it JWST} pointing. For comparison to our PISN rate, we include estimates from \citet{Mackey:03,Weinmann:05,Wise:05}; and \citet{Hummel:12}. {\it Bottom:} All-sky GRB rates per year. We here assume an ideal instrument with `perfect' sensitivity. For reference, we reproduce GRB rate estimates from \citet{Bromm:06,Campisi:11}. As can be seen, Pop~III GRBs are rare, and would require multi-year survey campaigns. 
}
\label{fig:trans}
\end{center}
\end{figure}

In the top panel of Figure~\ref{fig:trans}, we present the results for
Pop~III CCSN/PISN events (blue lines), and for Pop~II CCSNe (orange line), where the rate has been converted from per solid angle to per 10~arcmin$^{2}$.  The conversion is done to represent a {\it JWST} NIRCam pointing field of view. Our results imply that a future {\it JWST} survey, such as the 100 arcmin$^2$ CEERS program \citep{CEERS:18}, can expect event rates of $\sim 1\ {\rm yr^{-1}}$ for Pop~III CCSNe and $\sim 0.1\ {\rm yr^{-1}}$ for Pop~III PISNe, though multi-epoch follow-up would be needed to confirm any detection.  These results are roughly consistent with previous studies \citep[e.g.][]{Wise:05,Wiersma.etal:09,Hummel:12}. Differences are directly related to the underlying SFRDs, assumed in each study.

To estimate the GRB rate, we carry out a similar procedure as above, with the exception that $\zeta_{\rm IMF}$ is replaced with $\zeta_{\rm GRB}$, the GRB formation efficiency per unit mass, $\zeta_{\rm GRB}=2 \times 10^{-9}$. Furthermore, we now integrate over the entire sky ($4\pi$), thus removing the solid angle dependence, accounting for the fact that GRB detectors are not limited to observing a single patch of sky. From our Pop~II/III SFRDs, we estimate that a `perfect', all-sky instrument would find $\sim 0.1$ GRBs originating from Pop~III per year.

\section{Discussion of key topics}
\label{sec:disc}

\subsection{Comparison with previous studies}
\label{subsec:dev}
In Section~\ref{subsubsec:sfrd}, we made a detailed comparison between our simulated Pop~II SFRD and that from previous numerical works (see bottom panel of Fig.~\ref{fig:sfrd}). It is evident that our simulation is producing far more stars ($\sim 10$ times in some cases) at $z\gtrsim 10$. All numerical simulations include sophisticated models for star formation, cooling and feedback, with complex dependencies on one another. Therefore, a single simple cause for the deviation between our $z>10$ Pop~II SFRD and previous numerical predictions may not exist. In light of this, we discuss several scenarios which could contribute to the discrepancy.

{\it Scenario \#1:}  Our star formation routines (P3L, P2L) are simply producing too many stars.  This could be the case if our star formation efficiencies are too high ($\eta_{\rm *, III}=0.05$, $\eta_{\rm *, II}=0.10$).  For the case of Pop~II, direct observational estimates of these values at $z>6$ are currently not available, and thus we must rely on observations of local analogs for our simulated star forming regions.  The most appropriate of which would be giant molecular cloud (GMC) scale objects, with masses in the range $10^{3}\lesssim M/\Msun \lesssim 10^{5}$. GMCs have estimated star formation efficiencies in the range $\eta_{\rm *, II}\approx0.03-0.24$ \citep[e.g.][]{Kennicutt:98,Krumholz:07,Evans:09,Murray:11}, placing our value of $\eta_{\rm *, II}=0.10$ well within the observed range.

Potential overproduction of stars via our P3L model could indirectly lead to a corresponding overestimate for Pop~II stars, due to an unphysically rapid  metal enrichment of the primordial ISM. Admittedly, the Pop~III star formation efficiency is far more uncertain, as there are no direct observations or local analogs. Therefore, we rely on high-resolution, ab initio simulations of metal-free star forming regions to determine the mass of a single Pop~III stellar group, $M_{\rm *, III}\approx 500\Msun$, which is then distributed according to the IMF
\citep[e.g.][]{Greif:11,Hirano:14,Stacy:16}. Our efficiency factor of $\eta_{\rm *, III}=0.05$ is a consequence of this numerical calibration.

Our simulation could also overproduce stars if the stellar feedback prescription is too weak. Such feedback has long been recognized as a primary mechanism through which galaxies regulate their star formation \citep[e.g.][]{White:78, Dekel:86, White:91, Hopkins:12, Somerville:14}.  As the star formation life cycle progresses, the surrounding ISM receives large amounts of energy via stellar winds, radiation and SN shock fronts, removing entirely, or at least partially, gas which otherwise could have collapsed to form stars. Our P2L and P3L models are focused on the long-term legacy, left behind by early star formation, and they lack sufficient resolution to directly model feedback processes, such as radiation pressure, or expanding SN shock fronts.  However, our legacy models replicate key physical aspects of those processes, in that gas is heated, ionized, and the resulting overpressure moves gas from high-density star forming clouds to low-density ISM/IGM regions. It should be repeated that we do not artificially ``kick'' particles out of high density star forming regions via a sub-grid wind prescription.  Rather we rely on thermal energy injection and subsequent hydrodynamics to vacate these regions. This is a departure from previous methods and could result in artificially high star formation rates.  Future frontier observations will allow for better constraints and serve as a test for our approach.

While we feel justified in the adaptation of the physical parameters via observation and high resolution numerical experiments, the primary support for our models derives from the agreement with extrapolations from current observations of $z\gtrsim 6$ galaxies, in particular regarding the total predicted SFRD (Fig.~\ref{fig:sfrd}), SMF (Fig.~\ref{fig:smf}) and UVLF (Fig.~\ref{fig:uvlf}). Were our simulations dramatically overproducing stars, we would expect to see a strong departure from extrapolations of these robust observations. It is also worth noting that our work is consistent with the $z>6$ SFRD derived from GRB detections \citep[see ][]{Robertson:12,Yang:13}.

{\it Scenario \#2:}
Simulation resolution dictates the dynamic range which can be produced within a given volume.  For example, large cosmological volumes with box sizes $\sim 100h^{-1}\ {\rm Mpc}$, will contain objects at the massive/bright end of the SMF/UVLF, whereas volumes with sizes in the $10-50h^{-1}\ {\rm Mpc}$ range will reproduce the low mass/faint-end of those functions. The volume chosen for this work, with length $4h^{-1}\ {\rm Mpc}$, is specifically chosen to replicate a {\it JWST} deep-field pointing, with sufficient resolution elements to allow for pre-stellar clumps to be resolved. As a consequence, we are exploring the extreme low mass/faint-end of the SMF/UVLF, which is beyond the capabilities of {\it HST}.  With this in mind, we should not be surprised if our SFRD did not match with simulations designed to reproduce a different dynamic range in halo mass or designed to study a different epoch in cosmic evolution.

The situation is different for the FiBY simulation, analyzed in \citet{Johnson:13}, which has the same $4h^{-1}\ {\rm Mpc}$ box size, with more resolution elements ($684^3$), giving it slightly better mass and spatial resolution. Consequently, we are exploring a similar dynamic mass range. Yet, we are producing significantly more Pop~II stars over the redshift range $z=10-20$. We suspect that a stronger, `local' Lyman-Werner flux in the FiBY simulation could be the reason for the lower SFRD, via enhanced photo-dissociation of the low-temperature molecular coolants, $\Htwo$ and HD. We intend to explore this further in future work.


\subsection{To turn over or not to turn over}
\label{subsec:turn}
Additional factors to consider, when integrating an observed UVLF to produce a luminosity density or SFRD, are the limits of integration, and any deviation from the faint-end power-law slope, such as a turnover or flattening. Basic physical considerations suggest that the faint-end of the UVLF cannot continue  indefinitely towards ever fainter objects. At some point, it must turn over, or truncate. The existence and properties of this turnover have been the subject of a vigorous debate \citep[e.g.][]{Trenti:12, Jaacks:13, MBK:15, Livermore:17}.  
Previous numerical studies have predicted a turnover/flattening for $M_{\rm UV}\gtrsim -17$ at $z>6$ \citep{Jaacks:13, Wise:14, Oshea:15}. Observations of local dwarf galaxies suggest that a constant faint-end slope at $z\simeq 7$ of $\alpha_{\rm UV}= -2.0$, beyond $M_{\rm UV}\approx -13$, would result in $\sim 100$ times the number of dwarf galaxies than currently observed \citep{MBK:15}. Note, there is a high degree of uncertainty when trying to predict high-$z$ properties using local satellites.  However, HFF observations from \citet{Livermore:17} find no evidence for the faint-end slope deviating from the power-law predictions, at $z=6, 7, 8$, respectively, within the limiting magnitudes of $M_{\rm UV}=-12.5, -13.5, -15$ \citep[other analyses of this data agree that no turnover is present at $M_{\rm UV}<-15$;][]{Atek:15,Yue:16,Bouwens:17}.

In this work, we again find evidence for a flattening of the SMF and UVLF at $z= 8, 10, 15$ with $M_{\rm UV,turn}\approx -12, -13.5, -14$, respectively.  These results are in excellent agreement with those presented in \citet{Wise:14}, and with results in \citet{MBK:15}, who suggest that a flattening for $M_{\rm UV}\approx-13$ at $z$=7, with a subsequent slope of $\beta_{\rm UV}=-1.2$, is required to account for observations of local dwarfs. The UVLF turnover point, found here, lies just beyond the limiting magnitudes of \citet{Livermore:17}. This suggests that {\it JWST}, in conjunction with a lensing program similar to the HFF, could rule out or validate this key result. Furthermore, we find that this flattening corresponds well with the mass transition between atomic and molecular cooling haloes (see gray shaded region in the SMF, Fig.~\ref{fig:smf}). This result resonates with the idea that atomic cooling haloes support higher star formation efficiencies, due to their increased number of cooling channels \citep[atomic, molecular and metals;][]{Bromm:01a, Bromm:03, Santoro:06, Maio:10, Omukai:10}, and their ability to cool even in the presence of strong external radiation fields \citep[ionizing UV, soft LW;][]{Maio:07,Wise:07,CSS:10}. Therefore, we suggest that the star formation efficiency differential between atomic and molecular cooling haloes is the root cause of the flattening seen in this work.  

\begin{figure}
\begin{center}
\includegraphics[scale=0.40] {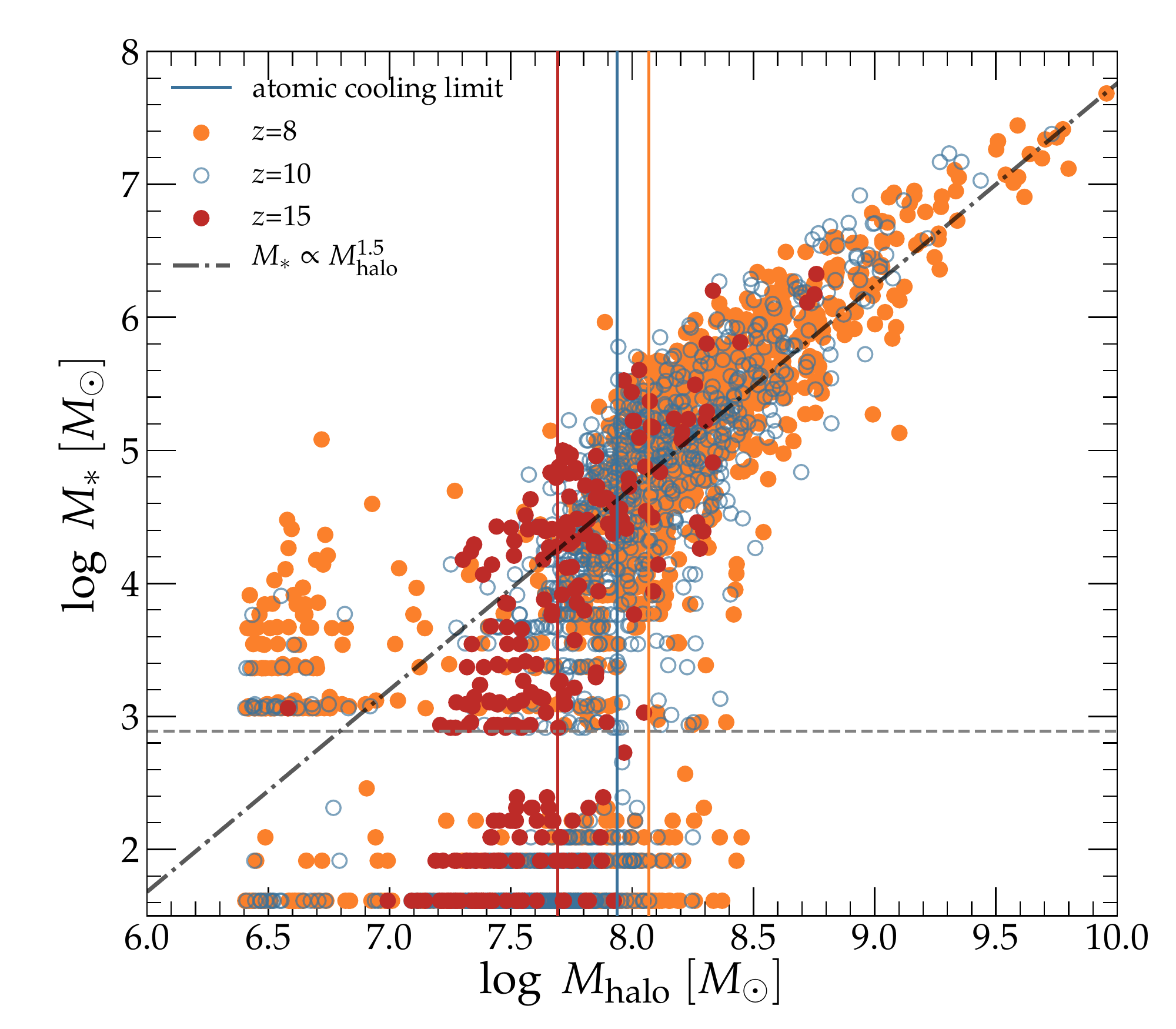}
\caption{Stellar mass to halo mass (SMHM) relation for $z=8, 10, 15$. The vertical lines correspond to the atomic cooling halo mass for each redshift.  Below the atomic cooling limit, we see a deviation from the $M_*\propto M_{\rm halo}^{1.5}$ power law, indicating a lower star formation efficiency in molecular cooling haloes. Note that the power law provides a fit to the $z=8$ data, for $M_{\rm halo}\geq 10^8\ \Msun$.
}
\label{fig:smhm}
\end{center}
\end{figure}

In the presence of strong UV background radiation, the accretion of IGM gas onto atomic cooling haloes can be suppressed, as any IGM gas heated to above the viral temperature of the halo will not accrete, resulting in lower star formation rates. Previous simulations have explored this Jeans-filtering process, and shown that it can take effect at halo masses above the atomic cooling limit, $M_{\rm halo}\lesssim 10^9-10 \Msun$ \citep[e.g.][]{Iliev:07, Mesinger:08, Okamoto.etal:08}, and has been suggested as a possible origin for a turnover in the UVLF. To explore this possibility, we consider the relationship between stellar mass and halo mass (SMHM). In Figure~\ref{fig:smhm}, we show the SMHM relation for $z=8, 10, 15$, along with vertical lines denoting the atomic cooling mass for each redshift.  If our simulated haloes were experiencing significant photo-suppression from UV background photons, we would expect to see a deviation in the slope of this relation at some mass above the atomic cooling limit. To the contrary, such deviation is not seen, although there is significant scatter for masses larger than the atomic cooling threshold. However, this is expected as in our simulation volume the UV background flux is still ramping up at $z$=8, with only $<50\%$ of the IGM ionized. Future work, continuing these simulations to lower redshift, can better assess the impact of photo-suppression on the turnover of the luminosity function at $z \lesssim 8$. It is worth noting that the absence of photo-suppression at these redshifts and halo masses is consistent with recent semi-analytic results presented in \citet{Yung:18}. 

\begin{figure}
\begin{center}
\includegraphics[scale=0.40] {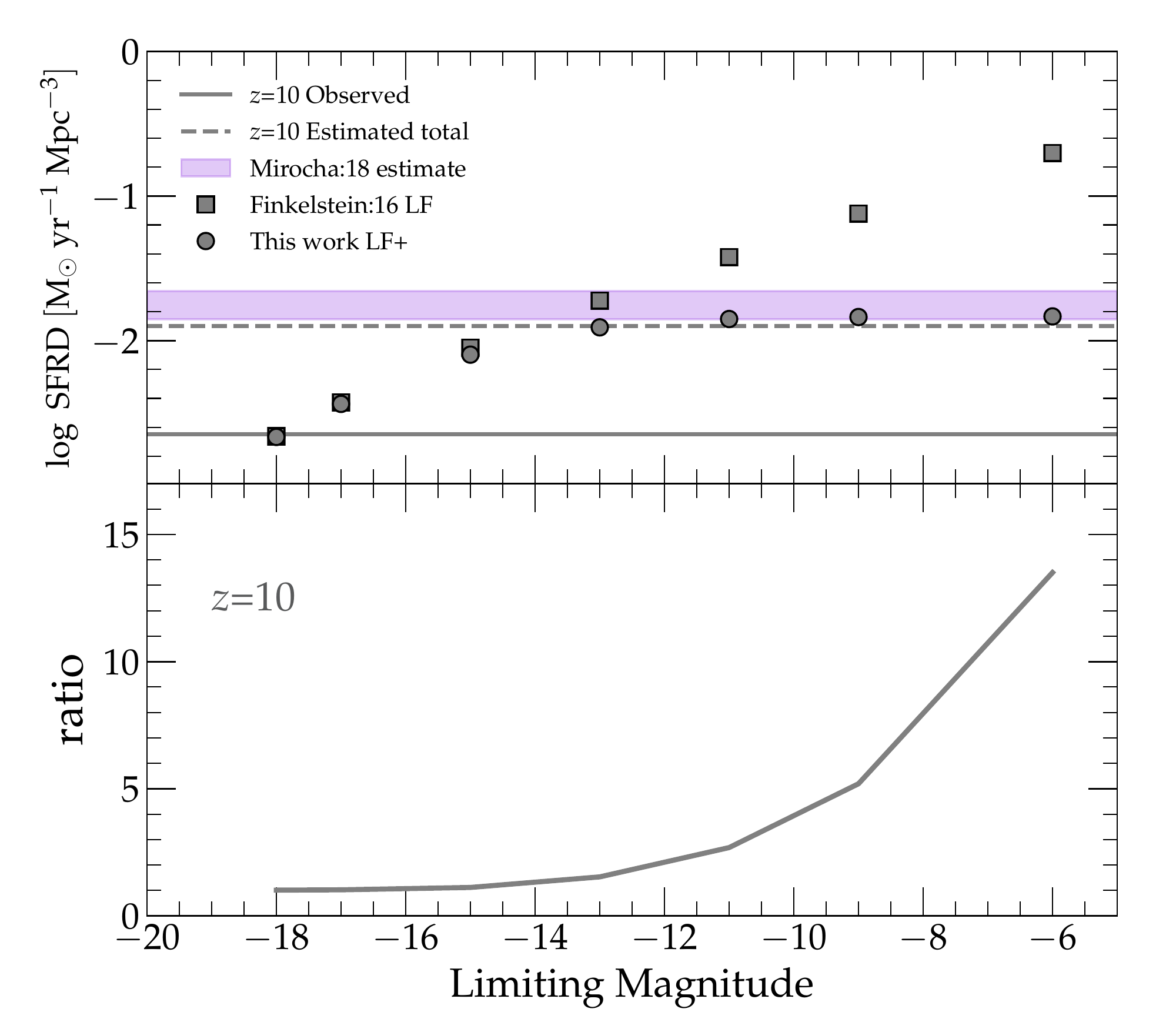}
\caption{Global impact of UVLF turnover. {\it Top:} Cosmic SFRD, derived by integrating the \citet{Finkelstein:16} $z=10$ UVLF to a range of limiting magnitudes (gray squares). For comparison, we show the same integration, using a functional form which includes a turnover in the faint-end power law \citep[gray circles;][]{Jaacks:13}. We also reproduce the $z=10$ estimate, derived from the EDGES constraint \citep[purple shaded region;][]{Mirocha:18}.
{\it Bottom:} Boost factor for SFRD in the absence of a turnover. Depending on the integration limiting magnitude, a single power-law UVLF fit, without a turnover, can result in a factor of $>10$ larger SFRD.
}
\label{fig:turn}
\end{center}
\end{figure}

A possible turnover in the UVLF would be reflected in estimates of the cosmic SFRD. To illustrate this effect, in Figure~\ref{fig:turn} we present the results from a numerical exercise, where we compare the SFRD at $z\simeq 10$, derived from a UVLF both with and without a turnover, employing the conversion from luminosity to stellar mass density in \citet{Madau:14}. The top panel shows the SFRD, derived with Schechter-fit parameters found in \citet{Finkelstein:16}, compared with values derived from the Schechter+ formulation \citep[see Equ.~\ref{eq:lf+};][]{Jaacks:13}. Note that we fix the standard UVLF Schechter parameters, in an effort to isolate the impact of including a broken power-law faint end ($\log \phi^*=-4.13,\ M_{\rm UV}=-20.25,\ \alpha_{\rm UV}=-2.35$). From this exercise, it is clear that, depending on the integration limiting magnitude, a single power-law UVLF fit can result in a factor of $>10$ difference in the estimated SFRD, if one integrates to the luminosity of a single O star. However, at a limiting magnitude of -13, assumed by most observational analyses, the difference is only $\sim10\%$.

It is interesting to note that, while our cosmic SFRD agrees very well with the estimates provided by \citet{Mirocha:18}, we arrive at this agreement with very different conclusions regarding the faint-end slope of the UVLF.  In contrast to our predicted UVLF flattening, their models require a steepening of the UVLF to be consistent with the EDGES measurements.  Future observations of the $z$>10 faint-end slope should help to differentiate between the diverging model predictions.

As was the case with results presented in \citet{Jaacks:13}, numerical resolution could play a role in identifying a possible UVLF turnover. To address this concern, we consider the study of \citet{Wise:14}, who find very similar results to ours. They use the grid-based AMR code {\small ENZO}, which is able to add additional levels of grid refinement to areas which require higher spatial resolution, such as regions of star formation. In the highest resolution zones, they achieve a $\sim 1$~(comoving)~pc grid size, and a dark matter particle mass of $1840\ \Msun$. This represents a much higher resolution than the work presented here (see Table~\ref{tbl:Sim}), yet we are finding extremely consistent results (see Fig.~\ref{fig:uvlf}). Therefore, we do not believe that the presences of a turnover in the UVLF is a consequence of insufficient numerical resolution. It is also worth noting that the numerical methodologies used in \cite{Jaacks:13}, \citet{Wise:14}, and here (i.e. {\small GADGET}, {\small ENZO}, and {\small GIZMO}), represent independent code development and verification streams. This provides additional confidence in the physical robustness of the turnover result.

\subsection{Termination of Pop~III}
\label{subsec:term}
A robust prediction from this work, as well as others \citep[e.g.][]{Johnson:13,Pallottini:15,Xu:16,Sarmento:18}, is that Pop~III star formation continues at a significant rate ($\sim 10^{-3}\ \Msun{\rm yr^{-1}\, Mpc^{-3}}$) at $z\lesssim 8$. While we do not extend our simulation to low enough redshifts to witness its termination, we know that it must end, based on the absence of metal-free star formation in local observations. There are three main mechanisms for terminating Pop~III star formation: (1) photo-dissociation of $\Htwo$/HD by LW radiation, (2) metal enrichment of all high-density gas to $Z>Z_{\rm crit}$, and (3) photo-ionization of primordial gas.  Even though we do not directly simulate the Pop~III termination here, we can place constraints on this process with lessons learned in this work.

Lyman-Werner (LW) photons, with energies in the range $11.2$ to $13.6$~eV, are able to destroy $\Htwo$/HD via photo-dissociation.  With the primary coolants thus destroyed, minihaloes will be unable to cool, collapse and form Pop~III stars. As detailed in \citet{Jaacks:17a}, we include a model for a global LW radiation background, tied directly to our simulated Pop~III/II SFRDs.  We also include a mechanism by which gas can self-shield at sufficient densities from the effects of an external LW radiation field \citep{Draine:96}.  While we notice that the LW background flux does act to suppress low-density gas from cooling, at high densities the self-shielding factor allows for Pop~III star formation to continue. This is consistent with high-resolution simulations which find that star formation can indeed continue in the presence of strong external radiation fields \citep{Maio:07,CSS:10}.

According to our model, Pop~III stars can only form in gas with a metallicity below $Z/\Zsun=10^{-4}=Z_{\rm crit}$.  Once this critical metallicity is reached for all gas in the simulation volume, Pop~III star formation will be terminated. However, from Figure~\ref{fig:vff} it is evident that, at $z$=7.5, only a small fraction of the gas in the simulation volume is enriched to beyond $Z_{\rm crit}$ ($\lesssim 10^{-3}$). This indicates that there is still a large reservoir of low-metallicity gas available to fuel ongoing Pop~III star formation, consistent with previous studies \citep[e.g.][]{Johnson:13,Muratov:13,Pallottini:15}.  

A strong UV background (UVB) can also photo-ionize and heat pristine gas, thereby suppressing star formation. As detailed in Section~\ref{subsubsec:heating}, we implement a model for a global UVB with photo-ionization rates adopted from \citet{Faucher.etal:09}. The ionization history implicit in this model assumes that the ionization fraction steeply increases, starting at $z\sim 10$, with the Universe being substantially ionized by $z$=6. Therefore, at $z$=7.5, when our simulation ends, the reionization process has only just begun. High-density gas is allowed to self-shield from the UVB, leaving a substantial amount of star-forming gas untouched. We have run low-resolution simulations to below $z\sim 6$, not shown here, to verify that the UVB does indeed suppress Pop~III star formation, once reionization is substantially complete.

In Section~\ref{subsubsec:heating}, we detail two sources which contribute to the total photo-heating experienced by a gas particle:  Internal heating ($\Gamma_{\rm pe}$) and external heating $\Gamma_{\rm uvb}$. Because the global UVB may contain flux from sources within our computational volume, there is the possibility of double counting photo-ionzing sources in our implementation of photo-heating.  However, as this work is focused on the pre-reionization Universe and the ionizing fraction at $z\sim7.5$ is still relatively low, our main results are unlikely to be affected by this potential problem. We defer further investigation of this effect to future work, when we plan to push our simulations to lower redshifts.

Based on the high Pop~III SFRD found in this work, extending down to $z\lesssim 8$, we suggest that the primary contributor to the termination of Pop~III is ultimately the ionizing UVB, which has yet to ramp up fully in our simulation. However, all three processes discussed above will likely play a role. We plan to quantify their relative contributions in future work.

\section{Summary and conclusions}
\label{sec:con}

Building upon our sub-grid model for Pop~III legacy star formation (P3L) presented in \citet{Jaacks:17a}, we implement a similar Pop~II legacy star formation module (P2L) for use in meso-scale cosmological volume simulations. With these star formation models, we study the metal-enrichment and evolution of galaxies in the first billion years of cosmic history by quantifying the individual Pop~III and Pop~II contributions. We analyze our simulation to make testable predictions for the upcoming {\it JWST} mission. Our major conclusions are as follows:

\begin{itemize}
\item We find that our Pop~III peak SFRD$\sim 10^{-3}\ \Msun{\rm yr^{-1}}$ is largely unchanged with the addition of Pop~II feedback physics, when compared to results in \citet{Jaacks:17a}, where metal-enrichment from Pop~II was not included. This suggests that continued Pop~III star formation is robust even in the presence of ongoing Pop~II star formation.

\item Our P2L star formation model provides excellent agreement (< factor of 2) with empirical constraints for the total cosmic SFRD at $z=7.5-10$.  At $z\gtrsim 10$, we find that our prediction falls within the upper and lower limits of the observation-based estimates (see top panel of Fig.~\ref{fig:sfrd}).

\item We find that the Pop~II SFRD quickly dominates Pop~III by $z\approx20$.  At $z=7.5$, Pop~III stars only contribute $\sim 5\%$ to the total cosmic SFRD. However, their high efficiency of producing ionizing photons allows them to contribute $\sim20\%$ to the total ionizing emissivity at $z=7.5$.

\item The number densities of our simulated galaxy populations show good agreement with both the observed stellar mass function (SMF), and the UV luminosity function (UVLF) at $z=8$ and $10$. Towards higher redshifts, at $z=15$, our results are consistent with observations, when extrapolating the faint-end of the observed UVLF with a fixed slope of $\alpha_{\rm UV}=-2.35$.

\item  Both our simulated SMF and UVLF show strong evidence of deviation from the low mass/faint-end power-law slope. This flattening occurs at $M_{\rm UV, turn}\approx -12, -13.5, -14$, for $z= 8, 10, 15$, respectively, and is strongly correlated with the mass transition between atomic cooling and molecular cooling haloes ($M_*\approx10^5\ \Msun$).  

\item We find that $\lesssim 10^{-3}$ of our simulation volume is enriched beyond $Z_{\rm crit}$ by $z$=7.5, leaving a large reservoir of pristine gas available to fuel ongoing Pop~III star formation. Furthermore, $\sim 20\%$ of Pop~III star formation at $z$=7.5 is occurring in isolated dark matter haloes which have experienced no Pop~II star formation events.

We are at a remarkable time in the history of astronomy, when we are just about to extend our empirical horizon to the epoch of cosmic dawn, when star and galaxy formation first began. Upcoming observations with frontier facilities, such as the {\it JWST} and the suite of extremely large ground-based telescopes, will reveal the crucial transition from initial cosmic simplicity to the proliferating complexity that ensued afterwards. To facilitate this grand observational program, numerical simulations are vital to provide predictions for the underlying model of $\Lambda$CDM cosmology. Our simulation presented here is a part of this endeavor, bringing into clearer focus the long-term legacy of cosmic star formation in the early Universe. 

There are a number of key lessons. On the one hand, star formation is able to establish pervasive radiation fields early on, in the form of soft (LW) UV radiation prior to reionization. Metal-enrichment from the first stellar generations, on the other hand, leaves behind a dual imprint. Locally, in the highly biased regions of the cosmic web, where star and galaxy formation is taking place, a significant `bedrock' level of metal enrichment ($\sim 1\%$~solar) is put in place. The sources seen by {\it JWST} in upcoming deep-field campaigns should thus all already be Pop~II systems, with Pop~III remaining largely hidden from view. This is an important test of the underlying bottom-up, hierarchical model of structure formation. Globally, in the low-density voids of the early IGM, most of the cosmic volume remains pristine. It is again an important challenge to test this prediction, with next-generation spectroscopic surveys of the diffuse IGM. Overall, we are getting closer to answering one of the fundamental questions in science: What are our cosmic origins, and how did it all begin?   

\end{itemize}
\section*{Acknowledgements}
This work was supported by HST-AR-14569.001-A \&  HST-AR-15028.001 (PI Jaacks), provided by NASA through a grant from the Space Telescope Science Institute, which is operated by the Association of Universities for Research in Astronomy, Inc., under NASA contract NAS5-26555. VB is supported by NSF grant AST-1413501. JJ and SLF acknowledge support from the NASA Astrophysics and Data Analysis Program award \#NNX16AN47G issued by JPL/Caltech. This work used the Extreme Science and Engineering Discovery Environment (XSEDE), which is supported by National Science Foundation grant number ACI-1548562, allocation number TG-AST120024.  The authors acknowledge the Texas Advanced Computing Center (TACC) at The University of Texas at Austin for providing HPC resources that have contributed to the research results reported within this paper.

\bibliographystyle{mnras}
\bibliography{jjref}
\bsp	
\label{lastpage}

\end{document}